\documentclass[10pt,letterpaper]{article}
\usepackage[top=0.85in,left=2.75in,footskip=0.75in,marginparwidth=2in]{geometry}

\usepackage{biblatex}
\usepackage{tikz}
\usetikzlibrary{
  positioning,
  calc,
  fit,
  backgrounds,
  arrows.meta,
  shapes.geometric
}

\usepackage{microtype}
\DisableLigatures[f]{encoding = *, family = * }

\raggedright
\usepackage{changepage}

\usepackage[aboveskip=1pt,labelfont=bf,labelsep=period,singlelinecheck=off]{caption}

\makeatletter
\renewcommand{\@biblabel}[1]{\quad#1.}
\makeatother

\usepackage{lastpage,fancyhdr,graphicx}
\fancyheadoffset[L]{2.25in}
\fancyfootoffset[L]{2.25in}

\usepackage{xcolor}
\definecolor{Gray}{gray}{.25}
\usepackage{tabularx}
\usepackage{multirow, multicol}
\usepackage{booktabs}
\usepackage{adjustbox}
\usepackage{float}

\usepackage{amsfonts, amssymb, amsmath, amsthm}
\DeclareMathOperator*{\argmax}{arg\,max}
\DeclareMathOperator*{\argmin}{arg\,min}

\newtheorem{theorem}{Theorem}
\newtheorem{proposition}{Proposition}
\newtheorem{remark}{Remark}
\newtheorem{lemma}{Lemma}

\usepackage{sidecap}
\usepackage{wrapfig}
\usepackage[pscoord]{eso-pic}
\usepackage[fulladjust]{marginnote}
\reversemarginpar

\usepackage{nameref,hyperref}

\begin{document}
\vspace*{0.35in}

\begin{flushleft}
{\Large
\textbf\newline{The Double Entropy Fit Index (DEFI): An Information-Theoretic Approach to Dimensionality Reduction and Clustering}
}
\newline
\\
Gabriel Lawrence\textsuperscript{1},
Daniel Otero-Le\'on\textsuperscript{2}
\\
\bigskip
\textbf{1} Department of Systems and Information Engineering, University of Virginia, \texttt{GabeL@virginia.edu}
\\
\textbf{2} Department of Systems and Information Engineering \& Department of Public Health Sciences, University of Virginia, \texttt{dfotero@virginia.edu}
\\
\end{flushleft}

\section*{Abstract}
We introduce the Double-Entropy Fit Index (DEFI), a regularized information-theoretic criterion for selecting among candidate graph partitions and reduced representations. DEFI couples entropy-based compression with relevance preservation and penalizes model dimensionality, cluster imbalance, and partition fragmentation. An exact R\'enyi-2 entropy identity shows that partitioning alters both spectral concentration and graph-volume normalization, explaining why unregularized entropy minimization can favor degenerate solutions. Across a 243-condition population-analytic factor recovery benchmark, DEFI achieved an overall exact recovery rate of 97.2\% (versus 75.0\% for EGA-TMFG, 45.8\% for TEFI on von Neumann entropy). In controlled local candidate batteries (distinguishing the true model from adjacent $K \pm 1$ and factor-split alternatives), DEFI identified the correct partition in 99.2\% of replications. Across ten clustering benchmarks, DEFI-guided Gaussian mixtures achieved a mean ARI of 0.929, exceeding HDBSCAN (0.809), Ward clustering (0.700), and conventional GMM criteria (0.501--0.580), with gains concentrating where standard criteria over-extract components.

\medskip
\noindent\textbf{Keywords:} Community Detection, Spectral Clustering, Dimensionality Reduction, Information Bottleneck, R\'enyi Entropy, Evolutionary Algorithms


\section{Introduction}
\label{sec:Intro}

Complex data in operations, engineering, and behavioral sciences are increasingly analyzed through the lens of network structure: variables or
observations are embedded in a graph, and the graph's communities are
interpreted as latent dimensions of the system. In psychometrics,
exploratory graph analysis (EGA) operationalizes this idea by treating
communities in a partial-correlation network as latent factors
\cite{golino2017exploratory}; the same reduction of a high-dimensional
system to a small number of internally coherent communities underlies
dimensionality reduction pipelines across machine learning and operations
research \cite{fortunato2010community, he2018learning}. In all of these
settings, a single question emerges: among candidate partitions of
the data, which one identifies the latent structure rather than an
artifact of the fit criterion?

This question is relevant because unregularized entropic criteria improve monotonically under partition refinement, favoring degenerate singletons or splinter clusters unless model complexity is priced explicitly \cite{jouvin2020high}. Existing remedies introduce additional operational constraints: network-based entropy criteria such as the von Neumann total entropy fit index (TEFI.vn) \cite{golino2021entropy} rely on inverse-covariance estimators \cite{JMLR:v9:drton08a, massara2016network} that become ill-conditioned under factor collinearity and are difficult to apply on disconnected or isolated-node graph structures, while classical mixture criteria (BIC, ICL) continue to reward spurious components on complex non-Gaussian geometries \cite{biernacki1998assessing}. 
What is missing is a unified criterion that scores candidate partitions directly from the spectrum of an affinity graph while explicitly pricing community count, cardinality imbalance, and topological fragmentation. We introduce the Double-Entropy Fit Index (DEFI), a unified partition-selection criterion designed to address this methodological gap. 

\subsection{Related Literature}
\label{sec:related}

Four frames of literature informed the work on our proposed Information-Theoretic selection criterion, DEFI. Entropy-based assessment of latent structure descends from Shannon's theory
of communication \cite{shannon1948mathematical} through Watanabe's total
correlation \cite{watanabe1960information}. For a collection of $n \ge 2$ observed 
random variables $X_1, \dots, X_n$ (e.g., manifest features or items), total 
correlation $C(X_1, \dots, X_n)$ generalizes bivariate mutual information as the 
gap between the sum of marginal entropies $H(X_i)$ and the joint entropy 
$H(X_1, \dots, X_n)$:
\begin{equation}
    C(X_1,\dots,X_n) = \sum_{i=1}^{n} H(X_i) - H(X_1,\dots,X_n).
    \label{eq:total-correlation}
\end{equation}
Additionally, Studen{\'y} and Vejnarov{\'a} showed that total correlation coincides
with the multiinformation function: it is nonnegative and vanishes exactly
under mutual independence \cite{studeny1998multiinformation}. Used directly as a partitioning objective,
total correlation is monotone under refinement and is minimized by the
trivial singleton partition; any usable criterion must therefore pair an
entropic score with control over the number and shape of communities
\cite{jouvin2020high, mcneish2018thorny}.

\paragraph{Partitioning as structure discovery.}
A second line of work makes the partition itself the estimand.
Compression-based formulations use minimum description length to choose the
number and composition of communities without prespecification
\cite{chakrabarti2004autopart}. Spectral methods translate community
detection into the eigenanalysis of Laplacian or modularity matrices
\cite{fortunato2010community, newman2013spectral}, and their detectability
limits under planted models are well characterized \cite{chen2015phase,
abbe2018}. In
psychometrics, exploratory graph analysis (EGA) estimates the number of
latent dimensions by community detection on partial-correlation networks
\cite{golino2017exploratory}, and the entropy fit indices built on this
machinery, most directly, the von Neumann total entropy fit index
(TEFI.vn) which scores candidate variable partitions by system entropy
\cite{golino2021entropy}; such network-based estimates compare favorably
with classical factor-retention rules in simulation
\cite{christensen2024comparing}. These criteria remain tied to
inverse-covariance network estimators \cite{JMLR:v9:drton08a,
massara2016network}, whose computational cost and connectedness
requirements are restrictive in high-dimensional or degenerate settings.

\paragraph{Entropy of the Laplacian state.}
DEFI computes entropy from the normalized combinatorial Laplacian
$\rho=L/\operatorname{Tr}(L)$, a density-matrix construction of
\cite{BraunsteinGhoshSeverini2006} whose R\'enyi generalization is
developed by \cite{DairykoEtAl2017}. Two findings from that literature
inform the design. First, the Laplacian-state entropy is dominated by
graph size and volume rather than by community structure, so it cannot be interpreted directly as a community descriptor
\cite{MinelloRossiTorsello2019}. Second, any partition-sensitive use of
the entropy must be explicit about normalization: the exact R\'enyi-2
change induced by deleting cut edges decomposes into a degree-sensitive
trace term and an opposing graph-volume term
(Section~\ref{sec:definitions-properties-assumptions}).

\paragraph{Regularization and the information bottleneck.}
DEFI's penalty grammar draws on classical model selection. The integrated completed likelihood augments a mixture likelihood with an entropy term that discourages degenerate assignments \cite{biernacki1998assessing}, and Bayesian Information Criterion (BIC) and Minimum Description Length (MDL) penalties price each additional community against sample size and dimension. The information-bottleneck principle formalizes the
trade-off between compressing an input and preserving information about a relevance variable \cite{tishby2000information, globerson2003sufficient,
shamir2010learning}; sufficient dimension reduction supplies the
corresponding statistical framework for information-preserving reduction
\cite{Adragni2009, Ghosh2022}.

\paragraph{Positioning.}
DEFI differs from this prior work in combination rather than in any single component. First, it evaluates a candidate partition through the R\'enyi entropy of the affinity-graph Laplacian state, which allows DEFI to avoid inverse-covariance estimation and tolerate disconnected graphs while pricing complexity, imbalance, and fragmentation explicitly. Second, it exposes one mixed-variable objective that a single derivative-free optimizer instantiates over mixture, hierarchical, and subspace candidate classes (Section~\ref{sec:definitions-properties-assumptions}). To the best of our knowledge, TEFI.vn is the closest published baseline and anchors our oracle and dimensionality comparisons.

\subsection{Contributions and Outline}
\label{sec:contributions}
This paper introduces the Double-Entropy Fit Index (DEFI), a regularized information-theoretic criterion for selecting among candidate partitions and reduced representations. We make four primary contributions: 
\textbf{(1)~Unified Information-Theoretic Criterion}: we formulate DEFI as a single objective coupling entropy-based compression, a self-supervised stability anchor via the deterministic information bottleneck, and structural regularizers for dimensionality, imbalance, and fragmentation, computed directly from the spectrum of an affinity Laplacian without inverse-covariance estimation (Section~\ref{sec:definitions-properties-assumptions}); 
\textbf{(2)~Exact Spectral Trace Decomposition}: we prove that at R\'enyi order $\alpha=2$, the change in spectral entropy decomposes into a trace-concentration term and an opposing graph-volume normalization term, providing the first analytical explanation for degenerate unregularized cuts and mathematically justifying DEFI's structural penalties (Section~\ref{sec:theory});
\textbf{(3)~Optimization Framework for Mixed Partition Spaces}: we solve the mixed discrete-continuous model-selection problem via continuous relaxation and CMA-ES, combining a piecewise linear homotopy over integer cluster counts with adaptive topological pre-conditioning and offline candidate pre-caching to eliminate step-discontinuity stagnation (Section~\ref{sec:CMA-ES-optimization}); and 
\textbf{(4)~Empirical Validation}: across a 243-condition factor-recovery benchmark, an exact enumeration across 700,074 structures, ten complex clustering geometries, and 26 high-dimensional manifold datasets, DEFI demonstrates superior recovery over classical information criteria, density methods, and factor retention rules (Sections~\ref{sec:evaluation-design}, \ref{sec:results}). Section~\ref{sec:discussion} contextualizes these findings within competing literature and details operational boundaries. Complete mathematical proofs, simulation derivations, and a comprehensive symbol index appear in the Electronic Companion (Sections~\ref{ec:identity}--\ref{ec:sec:notation}). Code and replication materials are publicly available at github will be available soon.

\section{The DEFI Criterion} 
\label{sec:definitions-properties-assumptions} 
\begin{figure}[!h]
\centering
\resizebox{0.9\textwidth}{!}{\input{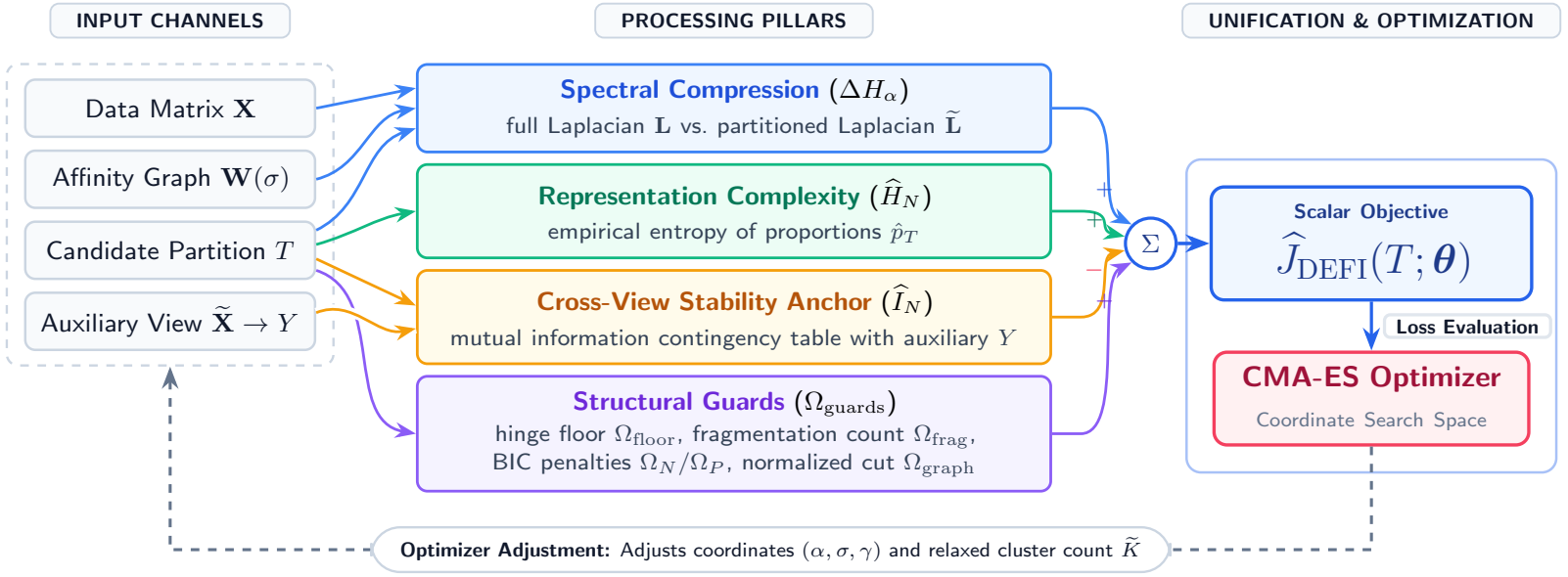}}
\caption{Operational architecture of the DEFI objective decomposition. Candidate partition $T$, affinity geometry $\mathbf{W}(\sigma)$, data matrix $\mathbf{X}$, and self-supervised reference view $Y$ map into four regularized processing pillars, aggregated into scalar objective $\widehat{J}_{\mathrm{DEFI}}(T;\boldsymbol{\theta})$ and optimized via CMA-ES coordinate feedback.}
\label{fig:defi-objective-architecture} 
\end{figure}

This section defines the Double-Entropy Fit Index (DEFI). A single principle organizes everything that follows: every experiment in this paper optimizes the same criterion, differing only in the candidate partition class $\mathcal{T}$ and free parameter space $\Theta$ (Table~\ref{tab:study-overview}). Conceptually, DEFI couples four foundational components: 
\begin{enumerate} 
\item[(i)] a spectral entropy term that measures affinity-graph concentration; 
\item[(ii)] a partition entropy term that prices representation complexity; 
\item[(iii)] a stability anchor derived from the deterministic information bottleneck; and 
\item[(iv)] structural regularizers that penalize degenerate, fragmented, or topologically discordant solutions. 
\end{enumerate} 
We formalize these components sequentially below (architectural overview in Figure~\ref{fig:defi-objective-architecture}), defining candidate partitions and the affinity-graph density state before assembling the global objective. A consolidated index of mathematical symbols, parameter domains, and operational definitions is cataloged in Section~\ref{ec:sec:notation} of the Electronic Companion. 

\subsection{Data, Partitions, and Candidate Classes} 
\label{sec:partitions} 
Let $\mathbf{X}=(\mathbf{x}_{1},\ldots,\mathbf{x}_{N})^{\top} \in\mathbb{R}^{N\times P}$ be a data matrix of $N$ observations and $P$ variables. A candidate partition is a deterministic assignment map 
\begin{equation} 
T:\{1,\ldots,N\}\rightarrow\{1,\ldots,K\}, 
\label{eq:partition-map} 
\end{equation} 
where $K$ is the candidate number of communities. For each community $c \in \{1, \dots, K\}$, its cardinality and proportion are given by: 
\begin{equation} 
n_c(T) = \sum_{i=1}^{N}\mathbb{I}(T_i=c), \qquad \widehat{p}_{T}(c) = \frac{n_c(T)}{N}, \qquad c = 1, \dots, K. 
\label{eq:community-proportions} 
\end{equation} 
When DEFI is applied to variables rather than observations (as in the factor-recovery studies), the domain of $T$ is $\{1,\ldots,P\}$, the denominator in \eqref{eq:community-proportions} is $P$, and the affinity graph is constructed on manifest features. A candidate class $\mathcal{T}$ denotes the family of admissible partitions produced by a given generator, and $\Theta$ is the bounded mixed discrete-continuous parameter space associated with that generator. In practice, $\mathcal{T}$ may be induced by parametric mixtures, cuts of an agglomerative dendrogram, or rotated subspace projections, with $\Theta$ capturing the corresponding structural and regularizer coordinates (Table~\ref{tab:study-overview}). 

\subsection{Graph Construction and the Laplacian State} 
\label{sec:graph-state} 
Each candidate configuration represents the data by a symmetric, nonnegative affinity matrix $\mathbf{W}=(W_{ij})\in\mathbb{R}^{N\times N}$ with $W_{ii}=0$. For a Gaussian-affinity graph, 
\begin{equation} 
W_{ij} = \exp\!\left(-\frac{\lVert\mathbf{x}_{i}-\mathbf{x}_{j}\rVert_{2}^{2}}{2\sigma^{2}}\right), \qquad i\neq j, 
\label{eq:affinity-definition} 
\end{equation} 
with bandwidth $\sigma>0$ scaled from the median positive pairwise distance; multiscale graphs evaluate convex combinations of affinities at bandwidths $0.5\sigma$, $\sigma$, and $2\sigma$. Let $\mathbf{D} = \operatorname{diag}(d_1,\ldots,d_N)$ with $d_i = \sum_{j} W_{ij}$ denote the degree matrix, and $\mathbf{L} = \mathbf{D} - \mathbf{W}$ the combinatorial Laplacian. Provided the graph contains at least one positive-weight edge, $\operatorname{Tr}(\mathbf{L}) > 0$ and the normalized density state $\boldsymbol{\rho} = \mathbf{L}/\operatorname{Tr}(\mathbf{L})$ is positive semidefinite with unit trace, forming a density matrix whose eigenvalues $\lambda_1,\ldots,\lambda_N$ are nonnegative and sum to one \cite{BraunsteinGhoshSeverini2006, DairykoEtAl2017}. Because only $\operatorname{Tr}(\mathbf{L})>0$ is required, the state remains well-defined on disconnected graphs and isolated vertices, unlike inverse-covariance network estimators. When an input similarity is not guaranteed positive semidefinite, DEFI projects it by clipping negative eigenvalues at zero and renormalizing to unit trace.

\subsection{Entropy Terms} 
\label{sec:entropy-terms} 
For a density matrix $\boldsymbol{\rho}$ and order $\alpha>0$, $\alpha\neq1$, the matrix R\'enyi entropy is 
\begin{equation} 
H_{\alpha}(\boldsymbol{\rho}) = \frac{1}{1-\alpha}\log\operatorname{Tr}(\boldsymbol{\rho}^{\alpha}) = \frac{1}{1-\alpha}\log\left(\sum_{i=1}^{N}\lambda_i^{\alpha}\right), 
\label{eq:matrix-renyi-entropy} 
\end{equation} 
whose limit as $\alpha\to1$ is the von Neumann entropy $H_{\mathrm{vN}}(\boldsymbol{\rho}) =-\operatorname{Tr}(\boldsymbol{\rho}\log\boldsymbol{\rho})$ ($0\log0:=0$). In CMA-ES search, $\alpha$ is optimized over $(0,1)\cup(1,100]$; candidates within $10^{-3}$ of $\alpha=1$ evaluate the von Neumann limit. The \emph{spectral entropy term} compares the graph state before and after the partition is imposed. Given $T$, let $\widetilde{\mathbf{L}}$ be the Laplacian obtained by deleting all inter-community edges, and let $\widetilde{\boldsymbol{\rho}}_{T} =\widetilde{\mathbf{L}}/\operatorname{Tr}(\widetilde{\mathbf{L}})$. Component~(i) evaluates the signed entropy difference 
\begin{equation} 
\Delta H_{\alpha}(T) = H_{\alpha}\!\left(\widetilde{\boldsymbol{\rho}}_{T}\right) - H_{\alpha}(\boldsymbol{\rho}). 
\label{eq:spectral-term} 
\end{equation} 
Its exact $\alpha=2$ decomposition, which motivates our structural regularizers, is derived in Section~\ref{sec:theory}. Component~(ii), the \emph{partition entropy}, prices representation complexity via the entropy of community proportions: 
\begin{equation} 
\widehat{H}_{N}(T) = -\sum_{c=1}^{K}\widehat{p}_{T}(c)\log\widehat{p}_{T}(c). 
\label{eq:partition-entropy} 
\end{equation} 

\subsection{The Relevance View and the Information Bottleneck} 
\label{sec:information_bottleneck} 
Component~(iii) grounds partition selection in the Information Bottleneck (IB) principle \cite{tishby2000information}, balancing compression against relevance preservation. In our unsupervised setting, the relevance variable $Y$ operationalizes a self-supervised cross-view stability anchor defined over the identical index set as $T$, derived from an auxiliary perturbation of the observed system. Let $M \in \{N, P\}$ denote the partitioned domain cardinality, corresponding to observation clustering ($M = N$) or variable partitioning ($M = P$). For observation clustering, an auxiliary view $\widetilde{\mathbf{X}} = \mathbf{X} + \boldsymbol{\mathcal{E}}$ is constructed via feature-scaled Gaussian perturbation ($\mathcal{E}_{ij} \sim \mathcal{N}(0, (0.1 \hat{\sigma}_j)^2)$, chosen to disrupt localized point density while preserving macroscopic cluster separation), where an initial unregularized baseline clustering on $\widetilde{\mathbf{X}}$ induces reference partition $Y \in \{1, \ldots, K_Y\}^N$ \cite{levine2001resampling}. For variable partitioning, an independently perturbed correlation matrix $\mathbf{R}_{\widetilde{\mathbf{X}}}$ yields reference assignment $Y \in \{1, \ldots, K_Y\}^P$. The reference anchor $Y$ is fixed prior to optimization and never derived from evaluation labels. Because candidate assignments are deterministic maps $T = f(\mathbf{X})$, assignments satisfy the Markov chain: 
\begin{equation} 
Y \leftrightarrow \mathbf{X} \leftrightarrow T. 
\label{eq:markovch} 
\end{equation} 
Mutual information between data and representation collapses to the partition entropy, $I(\mathbf{X}; T) = H(T)$, recovering the deterministic information bottleneck (DIB) \cite{shamir2010learning} where compression is governed directly by $\widehat{H}_M(T)$. For candidate partition $T \in \{1, \ldots, K\}^M$ and reference assignment $Y \in \{1, \ldots, K_Y\}^M$, the joint contingency is 
\begin{equation} 
\widehat{p}_{TY}(c, y) = \frac{1}{M} \sum_{i=1}^M \mathbb{I}(T_i = c, \, Y_i = y). 
\label{eq:contingency_joint} 
\end{equation} 
Preserved relevance is scored by plug-in mutual information ($0 \log(0/0) := 0$): 
\begin{equation} 
\widehat{I}_M(T; Y) = \sum_{c=1}^K \sum_{y=1}^{K_Y} \widehat{p}_{TY}(c, y) \log \frac{\widehat{p}_{TY}(c, y)}{\widehat{p}_T(c) \widehat{p}_Y(y)}. 
\label{eq:mutual_info_plugin} 
\end{equation} 
The signed term $-b \widehat{I}_M(T; Y)$ in the global objective rewards cluster boundaries that persist under stochastic jitter, penalizing noise overfit. 

\subsection{The Criterion and its Structural Regularizers} 
\label{sec:objective} 
Assembling components (i)--(iv) yields the DEFI criterion: 
\begin{equation} 
\begin{aligned} 
\widehat{J}_{\mathrm{DEFI}}(T;\boldsymbol{\theta}) ={} & a_{1}\,\Delta H_{\alpha}(T;\mathbf{W}) + a_{2}\,\widehat{H}_{N}(T) - b\,\widehat{I}_{N}(T;Y) \\ 
& + \gamma_{1}\,\Omega_{N}(K) + \gamma_{2}\,\Omega_{P}(K) \\ 
& + \gamma_{\mathrm{bal}}\,\Omega_{\mathrm{floor}}(T;\tau) + \gamma_{\mathrm{frag}}\,\Omega_{\mathrm{frag}}(T) \\ 
& + \eta\,\Omega_{\mathrm{graph}}(T;\mathbf{W}), 
\end{aligned} 
\label{eq:defi-objective} 
\end{equation} 
with nonnegative coefficients collected in $\boldsymbol{\theta}$ along with $\alpha$, graph parameters, and generator choices. The first three terms represent components (i)--(iii). The remaining terms formalize the structural regularizers of component~(iv): 

\paragraph{Dimensionality penalties.} BIC-type costs price candidate complexity against sample size and feature dimension: $\Omega_N(K) = K\log N$ and $\Omega_P(K) = K\log P$. 

\paragraph{Cluster balance floor.} Normalizing community-size entropy yields $\widetilde{H}_{\mathrm{size}}(T) = \widehat{H}_{N}(T)/\log K \in [0, 1]$, which attains 1 at uniform proportions. Given tolerance floor $\tau \in [0, 1]$, 
\begin{equation} 
\Omega_{\mathrm{floor}}(T;\tau) = \left[\max\!\left\{0,\, \tau-\widetilde{H}_{\mathrm{size}}(T)\right\}\right]^{2} 
\label{eq:entropy-floor-penalty} 
\end{equation} 
penalizes only severe imbalance without punishing substantively plausible asymmetry. 

\paragraph{Partition fragmentation penalty.} Splinter communities below a minimum cardinality $n_{\min}$ are priced by 
\begin{equation} 
\Omega_{\mathrm{frag}}(T) = \frac{1}{K}\sum_{c=1}^{K} \mathbb{I}\!\left\{n_c(T)<n_{\min}\right\}. 
\label{eq:fragmentation-penalty} 
\end{equation}

\paragraph{Graph topological concordance.} Disagreement with graph topology is penalized via the average normalized cut: 
\begin{equation} 
\Omega_{\mathrm{graph}}(T;\mathbf{W}) = \frac{1}{K}\sum_{c=1}^{K} \frac{\sum_{i\in V_c}\sum_{j\notin V_c}W_{ij}}{\sum_{i\in V_c}d_i}, 
\label{eq:graph-penalty} 
\end{equation} 
where $V_c=\{i:T_i=c\}$ and $\frac{0}{0}:=1$ maximally penalizes zero-volume sets. The selected partition solves 
\begin{equation} 
\widehat{T} \in \operatorname*{arg\,min}_{T\in\mathcal{T},\, \boldsymbol{\theta}\in\Theta} \widehat{J}_{\mathrm{DEFI}}(T;\boldsymbol{\theta}), 
\label{eq:defi-selected-partition} 
\end{equation} 
using CMA-ES (Section~\ref{sec:CMA-ES-optimization}). Because negative spectral differences $\Delta H_{\alpha}$ and relevance preservation $-b\widehat{I}_N$ can drive $\widehat{J}_{\mathrm{DEFI}}$ below zero while penalties $\Omega(T) \ge 0$, parameter space $\Theta$ is constrained to strictly positive penalty infima: $\Theta = \{ \boldsymbol{\theta} : \alpha \in (0, 1) \cup (1, 100], \, \sigma > 0, \, \gamma_k \in [\underline{\gamma}_k, \bar{\gamma}_k] \text{ with } \underline{\gamma}_k > 0 \}$. These positive floors ensure that search trajectories cannot evade structural regularization by collapsing weights to zero, forcing CMA-ES to adapt continuous kernel bandwidths and covariance geometries against active penalty boundaries.

\subsection{Properties and Assumptions}
\label{sec:properties}
Four structural properties govern this formulation: (i) only relative objective differences matter, so candidate rankings are invariant to constant additive shifts; (ii) all terms depend on $T$ strictly through the induced partition, ensuring invariance under community permutation; (iii) the balance regularizer is zero at uniform cluster proportions and strictly positive only below the floor $\tau$; and (iv) the signed spectral term $\Delta H_\alpha(T)$ requires structural regularization because, as proven in Section~\ref{sec:theory}, volume normalization can cause unregularized cuts to favor trivial partitions.

The implementation relies on four assumptions (theory-specific conditions for the internal-dominance branch and planted block models are stated with their respective theorems in Section~\ref{sec:theory}): 
(A1)~\emph{Well-defined graph state}: every decoded affinity graph contains at least one positive-weight edge and strictly positive bandwidths, with positive semidefinite graph states enforced via eigen-clipping; 
(A2)~\emph{Admissible partitions}: candidate partitions satisfy $2 \le K \le K_{\max}$ and the minimum-cluster cardinality constraint of \eqref{eq:fragmentation-penalty}, with numerically failed generator fits declared infeasible; 
(A3)~\emph{Nontrivial relevance view}: the reference configuration $Y$ is not almost surely identical to $\mathbf{X}$ and is independent of external evaluation labels; and 
(A4)~\emph{Estimator comparability}: all objective terms evaluate the identical observation set under uniform normalization conventions, and the parameter domain $\Theta$ is bounded.

\section{Theoretical Properties}
\label{sec:theory}

This section states the analytical properties of the DEFI criterion that
motivate its regularizers and justify its use as a selection rule.
Full proofs, and additional inferential results for the fixed-partition
coarse-graining appear in
the electronic companion (\ref{ec:identity}, \ref{ec:sbm},
\ref{ec:consistency}, \ref{ec:sec:restarts}). Throughout,
$\mathbf{W}\in\mathbb{R}^{N\times N}$ is a symmetric nonnegative
affinity matrix with zero diagonal and at least one positive-weight
edge, $\mathbf{L}$ its combinatorial Laplacian,
$V=\operatorname{Tr}(\mathbf{L})=\sum_i d_i>0$ its volume,
$S=\operatorname{Tr}(\mathbf{L}^2)=\sum_i d_i^2+\sum_{i\neq j}W_{ij}^2$,
and $\boldsymbol{\rho}=\mathbf{L}/V$ the graph state of
Section~\ref{sec:definitions-properties-assumptions}.

Our first result is the exact decomposition of the spectral term
\eqref{eq:spectral-term} at order $\alpha=2$. Given a partition $T$,
delete its inter-community edges, and for each vertex $i$ let
$c_i=\sum_{j:\{i,j\}\text{ cut}}W_{ij}$ be the weighted cut incidence.
Define the cut-volume and trace penalties
\begin{equation}
\begin{aligned}
    C(T) &= \sum_{i=1}^{N}c_i = 2\!\!\sum_{\{i,j\}\in\mathcal{E}_{\mathrm{cut}}(T)}\!\! W_{ij}, \\
    \varepsilon(T) &= \sum_{i=1}^{N}\bigl(2d_i c_i - c_i^2\bigr) + 2\!\!\sum_{\{i,j\}\in\mathcal{E}_{\mathrm{cut}}(T)}\!\! W_{ij}^2.
\end{aligned}
\label{eq:cut-penalties}
\end{equation}

\begin{theorem}[Exact R\'enyi-2 identity]
\label{thm:renyi2-identity}
Let $T$ be a partition whose deletion leaves at least one positive-weight
edge, with $0<C(T)<V$ and $0<\varepsilon(T)<S$. Then the spectral term
at $\alpha=2$ is exactly
\begin{equation}
    \Delta H_2(T)
    =
    -\log\!\left(1-\frac{\varepsilon(T)}{S}\right)
    +
    2\log\!\left(1-\frac{C(T)}{V}\right).
    \label{eq:renyi2-identity}
\end{equation}
For unweighted simple graphs ($W_{ij}\in\{0,1\}$),
$\varepsilon(T)=\sum_i c_i(2d_i-c_i+1)$.
\end{theorem}

The two terms in \eqref{eq:renyi2-identity} oppose each other: deleting
edges reduces $\operatorname{Tr}(\mathbf{L}^2)$ but also reduces the normalizing volume. Because $\boldsymbol{\rho}$ is normalized by
$\operatorname{Tr}(\mathbf{L})$, neither effect can be neglected, and the
second term rewards deletion per se. The consequence is a
failure mode for unregularized spectral-entropy minimization.

\begin{proposition}[Degeneracy of the unregularized criterion]
\label{prop:counterexample}
There exist connected graphs $G$ and partitions $T$ with no community
interpretation such that $\Delta H_2(T)<0=\Delta H_2(T_0)$, where $T_0$
is the no-cut partition. Hence minimizing \eqref{eq:spectral-term} alone
can strictly prefer a nontrivial partition over $T_0$, and minimizing it
over unrestricted families is not equivalent to minimizing the cut or
any degree-weighted cut. A three-vertex path with one end edge deleted
already exhibits the strict inequality.
\end{proposition}

Proposition~\ref{prop:counterexample} is the analytical reason
\eqref{eq:defi-objective} prices imbalance, fragmentation, and graph
disagreement rather than relying on the entropy difference alone. When
the volume effect is held fixed, the criterion does reduce to a
tractable combinatorial objective.

\begin{theorem}[Fixed-cut-volume equivalence] 
\label{thm:fixed-volume} 
On any admissible family $\mathfrak{P}_{C_0}=\{T:C(T)=C_0\}$ with $0\le C_0<V$, 
\begin{equation} 
\operatorname*{arg\,min}_{T\in\mathfrak{P}_{C_0}}\Delta H_2(T) \;=\; \operatorname*{arg\,min}_{T\in\mathfrak{P}_{C_0}}\varepsilon(T). 
\end{equation} 
Because $u \mapsto -\log(1-u/S)$ is strictly increasing on $[0,S)$, minimizing $\Delta H_2$ at fixed volume is equivalent to minimizing $\varepsilon(T)$. The induced penalty $\varepsilon(T)$ couples cut edges sharing an incident node via the negative quadratic term $-\sum_i c_i^2$, distinguishing it from a standard linear minimum cut. 
\end{theorem}

The next results interpret the criterion under a planted model. The
scope here is narrow: the analysis concerns a
\emph{fixed} partition and the binary edge-mass coarse-graining induced
by it.

\begin{theorem}[Population bound under the equal-sized Stochastic Block Model]
\label{thm:sbm-population}
Let $G$ be generated by the symmetric $q$-community planted-partition
model with equal community sizes $s$, internal probability
$p_{\mathrm{in}}$, external probability $p_{\mathrm{out}}$, and
$r_{q,s}=(q-1)s/(s-1)$ external-to-internal edge-opportunity ratio. Let
$\pi=p_{\mathrm{in}}M_{\mathrm{in}}/
(p_{\mathrm{in}}M_{\mathrm{in}}+p_{\mathrm{out}}M_{\mathrm{out}})$
be the population internal-edge mass, where
$M_{\mathrm{in}}=q\binom{s}{2}$ and
$M_{\mathrm{out}}=\binom{q}{2}s^2$. For $\alpha > 1$, on the internal-dominance branch $\pi \ge \tfrac{1}{2}$ (equivalent to $p_{\mathrm{in}}/p_{\mathrm{out}} \ge r_{q,s}$), and for any $\epsilon \in [0, \log 2]$,
\begin{equation}
    h_{\alpha}(\pi)\le\epsilon
    \quad\Longleftrightarrow\quad
    \frac{p_{\mathrm{in}}}{p_{\mathrm{out}}}
    \;\ge\;
    r_{q,s}\,\frac{B_{\alpha}(\epsilon)}{1-B_{\alpha}(\epsilon)},
    \label{eq:sbm-bound}
\end{equation}
where $h_{\alpha}(x)=(1-\alpha)^{-1}\log(x^{\alpha}+(1-x)^{\alpha})$ is
the binary R\'enyi entropy and $B_{\alpha}(\epsilon)$ is its unique
solution of $h_{\alpha}(B)=\epsilon$ on $[\tfrac12,1]$. Without the
branch condition, low entropy is symmetric between internal and
external dominance and \eqref{eq:sbm-bound} does not follow.
\end{theorem}

In words, Theorem~\ref{thm:sbm-population} defines the sharp signal-to-noise ratio ($p_{\mathrm{in}}/p_{\mathrm{out}}$) required for the planted partition to achieve an entropy deficit $\epsilon$, proving that low spectral entropy guarantees cohesive community structure only along the internal-dominance branch.

\begin{proposition}[R\'enyi--2 Bridge under the Planted Partition Model] 
\label{prop:renyi2_bridge} 
Let $G_n$ be an unweighted simple graph generated from the symmetric $q$-community stochastic block model (SBM) of Theorem~\ref{thm:sbm-population}, and let $T^\ast$ denote the planted partition. Define the empirical internal-edge mass and trace-penalty ratio by
\[
\pi_n = \frac{X_{\mathrm{in}}}{X_{\mathrm{in}} + X_{\mathrm{out}}}, \qquad \eta_n = \frac{\epsilon(T^\ast)}{S}.
\]
Then the spectral entropy difference at order $\alpha=2$ satisfies the exact finite-sample identity
\[
\Delta H_2(T^\ast) = 2\log \pi_n - \log(1 - \eta_n).
\]
Moreover, in the dense regime with fixed $p_{\mathrm{in}}, p_{\mathrm{out}} \in (0, 1)$, $(\pi_n, \eta_n) \xrightarrow{p} (\pi, \eta)$ where $1 - \eta = \pi^2 + O(n^{-1})$. Consequently,
\[
\Delta H_2(T^\ast) = o_{\mathbb{P}}(1) \quad \text{as } n \to \infty.
\]
\end{proposition}

\begin{remark}[Role of the Spectral Term and Scale Calibration]
\label{rem:spectral_calibration}
Proposition~\ref{prop:renyi2_bridge} bridges the macroscopic edge-mass statistics of Theorems~\ref{thm:sbm-population} and~\ref{thm:sbm-clt} with the spectral criterion of Theorem~\ref{thm:renyi2-identity}. Because $1 - C(T^*)/V = \pi_n$ holds identically on unweighted simple graphs, the volume-normalization term in $\Delta H_2$ is governed directly by the internal-edge mass $\pi_n$. 

The asymptotic vanishing $\Delta H_2(T^\ast) = o_{\mathbb{P}}(1)$ reflects the calibrated null behavior of the criterion evaluated strictly at the planted partition $T^\ast$. Across the broader candidate space $\mathcal{T}$, candidate partitions that violate true community boundaries yield $1 - \eta(T) \ne \pi(T)^2$, producing non-vanishing spectral penalties that separate cohesive structures from arbitrary cuts. Furthermore, while the combinatorial regularizers ($\Omega_N, \Omega_{\mathrm{floor}}, \Omega_{\mathrm{frag}}$) depend solely on cluster cardinalities, $\Delta H_\alpha(T; \mathbf{W})$ serves as the primary coupling between the partition and the continuous affinity geometry $\mathbf{W}$. Finally, the asymptotic cancellation $2\log\pi - \log(1-\eta) \to 0$ ensures that the spectral term operates on the same effective $\mathcal{O}(1)$ scale as the information-theoretic and structural regularizers. This stabilizes the objective where unregularized spectral differences flatten, preventing graph volume from overwhelming model-order selection while formally justifying the structural regularizers in \eqref{eq:defi-objective}.
\end{remark}

\begin{theorem}[Asymptotic distribution of the empirical entropy]
\label{thm:sbm-clt}
Let $\widehat{P}_n=X_{\mathrm{in}}/(X_{\mathrm{in}}+X_{\mathrm{out}})$
be the empirical internal-edge mass in the model of
Theorem~\ref{thm:sbm-population}, with
$X_{\mathrm{in}}\sim\mathrm{Bin}(M_{\mathrm{in}},p_{\mathrm{in}})$ and
$X_{\mathrm{out}}\sim\mathrm{Bin}(M_{\mathrm{out}},p_{\mathrm{out}})$
independent. If both count variances diverge and $\pi_n$ is bounded away from $0$ and $1$, then $\sqrt{V_{P,n}^{-1}}(\widehat{P}_n-\pi_n) \xrightarrow{d} \mathcal{N}(0,1)$ for the explicit variance $V_{P,n}$ derived in \ref{ec:sbm}. Consequently: (i)~if $\pi_n \to \pi \neq \tfrac12$, the plug-in entropy $h_{\alpha}(\widehat{P}_n)$ is asymptotically normal with mean $h_{\alpha}(\pi_n)$ and variance $[h_{\alpha}'(\pi_n)]^2 V_{P,n}$; (ii)~if $\pi_n = \tfrac12$ eventually, the first-order limit degenerates to the second-order $\chi_1^2$ distribution, $(h_{\alpha}(\widehat{P}_n) - \log 2)/V_{P,n} \xrightarrow{d} -2\alpha\,\chi_1^2$; and (iii)~in the dense regime $\widehat{P}_n - \pi_n = \mathcal{O}_{\mathbb{P}}(n^{-1})$, whereas under sparse scaling ($p_{\mathrm{in}}=a/n, p_{\mathrm{out}}=b/n$) the convergence rate slows to $\mathcal{O}_{\mathbb{P}}(n^{-1/2})$.
The balanced point $\pi=\tfrac12$ is simultaneously the entropy maximum
and the branch boundary, which is where the second-order limit applies. 
These statements concern
inference for a fixed partition; if the same graph is used both to
select a partition and to estimate its entropy, selection effects
invalidate the fixed-partition variances.
\end{theorem}

The criterion \eqref{eq:defi-objective} is a plug-in estimator
of the population regularized deterministic information-bottleneck
objective
\begin{equation}
    \mathcal{J}_{\mathrm{RDIB}}(T)
    =
    a_2\,H(T)-b\,I(T;Y)+\Omega_{\gamma}(T),
    \label{eq:population-rdib}
\end{equation}
where the compression term $I(\mathbf{X}; T)$ has been replaced by $H(T)$ via the Markov chain condition in \eqref{eq:markovch}, and the spectral and structural terms are collected in $\Omega_{\gamma}(T)$.

\begin{theorem}[Selection consistency over finite classes]
\label{thm:consistency}
Suppose that for every fixed admissible $T$ the entropy, relevance, and
regularizer estimators in \eqref{eq:defi-objective} converge in
probability to their population counterparts. Then
$\widehat{J}_{\mathrm{DEFI}}(T;\boldsymbol{\theta})
\xrightarrow{\mathbb{P}}\mathcal{J}_{\mathrm{RDIB}}(T)$ pointwise.
If, in addition, the candidate class $\mathcal{T}\times\Theta$ is finite
and the population minimizer $T^\star$ is unique and separated by a
margin $\Delta>0$, then the empirical minimizer satisfies
$\mathbb{P}(\widehat{T}_N=T^\star)\to1$. In particular, for fixed $P$,
fixed $K_{\max}$, and a discretized $\Theta$, the full class of
partitions into at most $K_{\max}$ communities is finite, so the
consistency statement covers the entire partition class, not only a
hand-picked candidate set.
\end{theorem}

Two scope statements delimit Theorem~\ref{thm:consistency}. First, with
a continuous $\Theta$, uniformity requires a uniform law of large
numbers for the entropy and mutual-information estimators over $\Theta$,
which we do not establish; the exact-enumeration study of
Section~\ref{sec:results} provides computational evidence for the
discretized case. Second, consistency identifies the population
minimizer of the regularized objective.

\begin{remark}[Theoretical Scope and Boundaries]
\label{rem:theory_scope}
To prevent conflation between structural properties and generative recovery guarantees, we delineate three precise boundaries of the theoretical analysis: 
(i)~\emph{Structural vs.\ Generative}: Theorems~\ref{thm:renyi2-identity} and~\ref{thm:fixed-volume} provide structural decompositions of the R\'enyi-2 spectral term on arbitrary graphs, identifying the volume-normalization artifact that necessitates structural regularization, rather than establishing minimax recovery rates under planted random graph models; 
(ii)~\emph{Calibration vs.\ Detectability}: Proposition~\ref{prop:renyi2_bridge} establishes asymptotic null calibration ($\Delta H_2(T^*) = o_{\mathbb{P}}(1)$) at the true partition under dense block models, ensuring scale compatibility with complexity penalties rather than bounding sharp information-theoretic detectability thresholds; and 
(iii)~\emph{Finite-Class vs.\ Model-Order Consistency}: Theorem~\ref{thm:consistency} guarantees selection consistency of the DEFI minimizer over finite, margin-separated candidate classes $\mathcal{T} \times \Theta$, but does not constitute an unconstrained proof of model-order consistency ($\mathbb{P}(\widehat{K} = K^*) \to 1$) over unbounded, continuous partition spaces.
The empirical studies in Section \ref{sec:results} 
evaluate the operational recovery performance of the full pipeline where exact 
high-dimensional asymptotic analysis remains an open problem.
\end{remark}

\section{Optimization with the Covariance Matrix Adaptation Evolution Strategy}
\label{sec:CMA-ES-optimization}

The model selection problem \eqref{eq:defi-selected-partition} couples discrete decisions ($K \in \{2, \dots, K_{\max}\}$, covariance family $m_{\mathrm{cov}}$) with continuous coordinates ($\alpha, \boldsymbol{\sigma}, \boldsymbol{\gamma}$). We embed these parameters into a bounded 15-dimensional surrogate domain $\widetilde{\Theta} \subset \mathbb{R}^{15}$ solved via CMA-ES \cite{hansen2006cma} (parameter bounds and encodings in Section~\ref{ec:sec:restarts}). To eliminate discrete step discontinuities, cluster counts are relaxed to real coordinates $\tilde{K} \in [2, K_{\max}]$, bounding integer floors $k_{\mathrm{fl}} = \lfloor \tilde{K} \rfloor$, ceilings $k_{\mathrm{ce}} = \lceil \tilde{K} \rceil$, and fractional weight $w_k = \tilde{K} - k_{\mathrm{fl}} \in [0, 1)$. Denoting bounding partitions by $T_{k_{\mathrm{fl}}}$ and $T_{k_{\mathrm{ce}}}$ under continuous covariance coordinate $m_{\mathrm{cov}} = \mathcal{M}[\operatorname{clamp}(\operatorname{round}(\tilde{m}), 1, |\mathcal{M}|)]$, the objective is evaluated as the linear homotopy:
\begin{equation}
    \widetilde{J}_{\mathrm{DEFI}}(\tilde{\boldsymbol{\theta}}) = (1 - w_k)\,\widehat{J}_{\mathrm{DEFI}}\big(T_{k_{\mathrm{fl}}}; \operatorname{decode}(\tilde{\boldsymbol{\theta}})\big) + w_k\,\widehat{J}_{\mathrm{DEFI}}\big(T_{k_{\mathrm{ce}}}; \operatorname{decode}(\tilde{\boldsymbol{\theta}})\big).
    \label{eq:smooth_k_interpolation}
\end{equation}
When $\tilde{K}$ is an integer ($w_k = 0$), \eqref{eq:smooth_k_interpolation} recovers DEFI identically. Between integer boundaries, \eqref{eq:smooth_k_interpolation} creates a continuous descent surface that prevents premature step-size deflation, while proposals with cluster sizes $n_c < 2$ receive an infeasibility penalty $P_{\mathrm{inf}} = +10^9$.

\begin{figure}[!ht]
\centering
\includegraphics[width=0.75\textwidth]{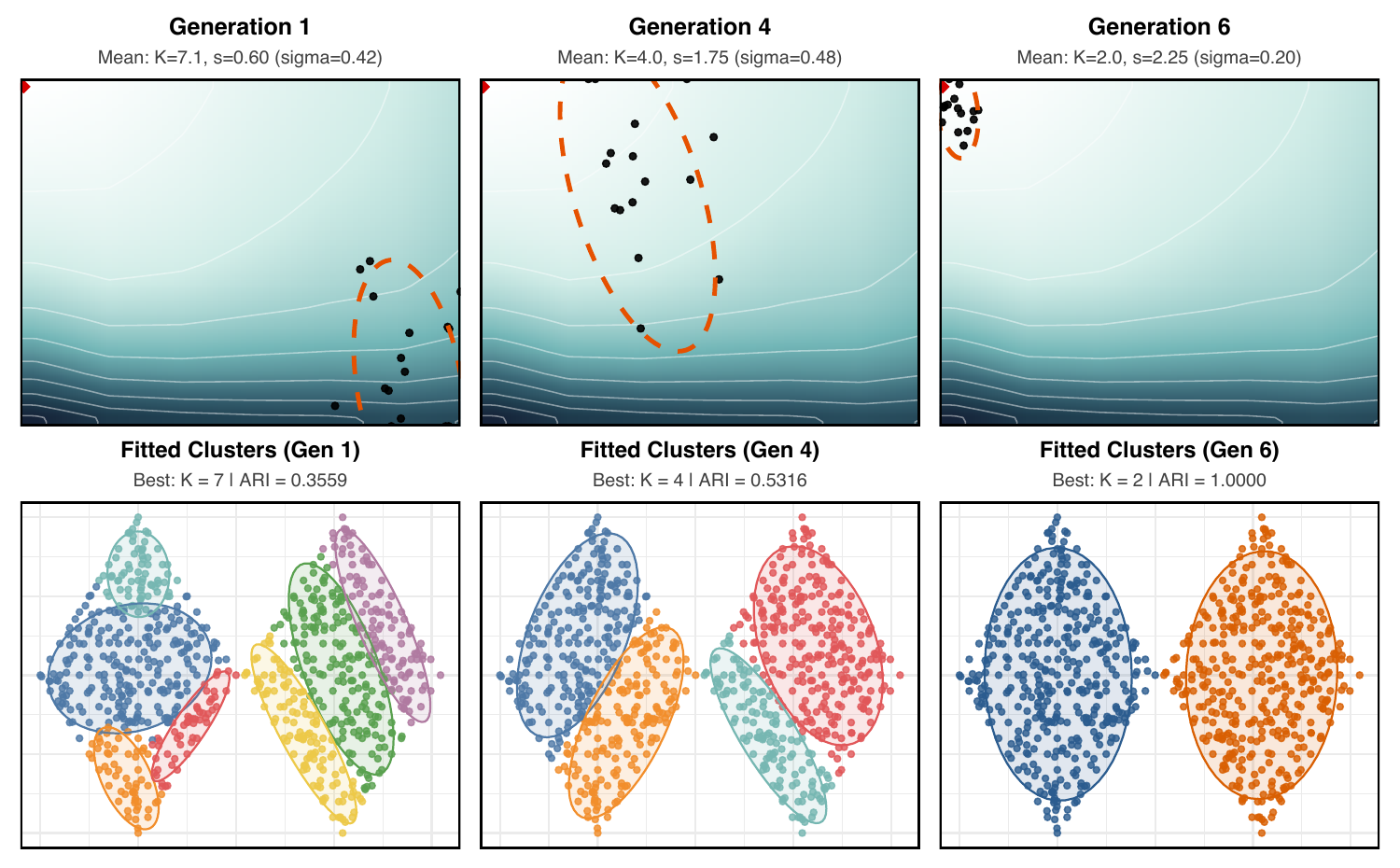}
\caption{CMA-ES optimization dynamics under continuous relaxation on the \emph{TwoDiamonds} topology ($b = 300$). 
\textbf{Top row:} Evolution of the sampling distribution mean $\mathbf{m}^{(g)}$, adaptive covariance ellipse $\mathbf{C}^{(g)}$ (dashed orange), and offspring proposals (black points) across the continuous surrogate landscape $\widetilde{\Theta}$. 
\textbf{Bottom row:} Corresponding decoded cluster partitions $T_K$ and covariance geometries in sample space. 
The linear homotopy \eqref{eq:smooth_k_interpolation} provides continuous descent directions that steer the population from initial over-partitioning ($\tilde{K} \approx 7.1$, ARI $= 0.356$) to basin convergence ($\widehat{K} = 2$, ARI $ = 1.000$) without integer step-discontinuity stagnation.}
\label{fig:cmaes-homotopy-trajectory}
\end{figure}

Figure~\ref{fig:cmaes-homotopy-trajectory} illustrates this search trajectory on the \emph{TwoDiamonds} topology. Even when initialized in an over-partitioned regime ($K \approx 7$), the adaptive covariance matrix $\mathbf{C}^{(g)}$ contracts along continuous homotopy slopes, systematically shedding spurious sub-clusters until converging onto the ground truth ($K=2$, $\text{ARI}=1.000$).

To prevent CMA-ES from allocating evaluation budget to geometrically incompatible covariance structures, an exploratory topological screening phase pre-conditions the search space before optimization begins. Point densities $\hat{\rho}_i$ estimated from local $k$-NN distances first yield a density-contrast ratio, $\kappa_{\mathrm{dens}} = Q_{0.95}(\{\hat{\rho}_i\}) / Q_{0.05}(\{\hat{\rho}_i\})$. Pairing $\kappa_{\mathrm{dens}}$ with an exploratory unregularized mixture fit classifies the data into one of three structural regimes (asymmetric manifold, equal-volume manifold, or standard heterogeneous), which initializes $\tilde{K}_0$ and prunes inadmissible covariance parameterizations from the candidate pool $\mathcal{M}$ (decision rules detailed in Section~\ref{ec:sec:restarts}). Finally, this screening step calibrates surrogate-domain boundaries by enforcing $\gamma_{\mathrm{bal}} \ge 0.15$ and bounding the dimensionality-penalty ceiling by $\bar{\gamma}_2 = \max(2.0, \operatorname{round}(c \cdot N / \log N))$.

Computational efficiency is achieved by decoupling candidate generation from fitness evaluation. In an offline phase, Gaussian mixtures are fitted once across candidate configurations $(K, m)$ via Expectation-Maximization and cached in an in-memory table $\mathcal{M}_{\mathrm{cache}}$. During CMA-ES search, evaluating \eqref{eq:smooth_k_interpolation} requires an $\mathcal{O}(1)$ partition retrieval from $\mathcal{M}_{\mathrm{cache}}$ followed by compiled C++ trace evaluations via \texttt{RcppArmadillo}. By exploiting the exact R\'enyi-2 trace identity of Theorem~\ref{thm:renyi2-identity}, evaluating $C(T)$ and $\varepsilon(T)$ requires only sparse edge-cut traversals scaling as $\mathcal{O}(|\mathcal{E}_{\mathrm{cut}}(T)|)$ rather than dense matrix operations. Consequently, uncached evaluations execute in under $60\,\mu\mathrm{s}$, completing the entire 300-evaluation CMA-ES run within seconds.

\section{Evaluation Design}
\label{sec:evaluation-design}

The evaluation is organized so that each study isolates a unique property of DEFIdistinct question. 
Study~1 removes search entirely and evaluates candidate structures directly at fixed 
default weights across both a 243-condition population-analytic factorial grid and the 
complete partition lattice of small instances. Study~2 attaches the CMA-ES optimizer 
and asks whether the full pipeline is competitive on observation-level clustering across 
heterogeneous benchmark topologies. Study~3 evaluates intrinsic dimensionality selection 
and non-linear manifold unfolding on high-dimensional benchmarks. Finally, Study~4 
conducts a mechanistic ablation isolating the contributions of the criterion, continuous 
relaxation smoothing, and covariance adaptation dynamics.

\begin{table}[!ht]
\begin{adjustwidth}{-1.75in}{0in}
\centering
\small
\caption{Evaluation Design Matrix: Experimental protocols, search configurations, relevance anchors, and evaluation endpoints across Studies~1--4.}
\label{tab:study-overview}
\begin{adjustbox}{max width=\linewidth}
\begin{tabular}{l llllll}
\toprule
\textbf{Study / Focus} & \textbf{Data \& Suite} & \textbf{DEFI Variant} & \textbf{Optimizer \& Budget} & \textbf{Relevance Anchor $Y$} & \textbf{Primary Endpoints} & \textbf{Baselines} \\
\midrule
\textbf{Study 1a} & 243-condition latent & DEFI-FA & Frozen defaults & Perturbed correlation & Exact recovery (\%) & EGA-TMFG, PA, Kaiser, \\
Factor Recovery & factor grid ($R = 100$) & & (no search; Sec.~\ref{sec:definitions-properties-assumptions}) & matrix ($M = P$) & $\Delta K = \widehat{K} - K^\dagger$ & TEFI.vn, SEM fit (CFI/TLI) \\
\addlinespace
\textbf{Study 1b} & Stirling partition & DEFI-FA & Exact enumeration & Perturbed correlation & Global argmin & Exhaustive search over all possible  \\
Finite Consistency & lattice ($P \le 12$) & & ($\le 700{,}074$ structures) & matrix ($M = P$) & recovery (\%) & graphs based on the Stirling numbers \\
\addlinespace
\textbf{Study 2} & 10 FCPS \& synthetic & DEFI-GMM & CMA-ES continuous & Feature Gaussian & Adjusted Rand Index, & GMM (BIC \cite{biernacki1998assessing}, ICL \cite{scrucca2016mclust}), HDBSCAN \cite{stewart2022implementation}, \\
Clustering & benchmarks (20 seeds) & DEFI-HC & relaxation ($b = 300$) & jitter view ($M = N$) & $\Delta K$, NMI, V-measure & Ward \cite{ward1963hierarchical}, $k$-means (silhouette) \cite{punhani2022binning} \\
\addlinespace
\textbf{Study 3} & 26 UCI benchmarks & DEFI-DR & Dual-projection rank & Perturbed correlation & $|\Delta m| = |\widehat{m} - K^\dagger|$, & Isomap, UMAP, Kernel PCA, \\
Dimensionality & \& MNIST (20 seeds) & & sweep ($M \le P$) & matrix ($M = P$) & Geo.\ stress, Var.\ (\%) & PA, Kaiser, Sparse PCA, VAE \\
\addlinespace
\textbf{Study 4} & 6 representative & DEFI-GMM & Controlled ablation & Feature Gaussian & ARI, $\Delta K$, Oracle & DEFI-Fixed, Random search, \\
Ablation & topologies (20 seeds) & & ($b = 300$ budget) & jitter view ($M = N$) & gap $\widehat{G}$ ($\epsilon \in \{0, 0.05\}$) & DEFI (no smooth-$K$) \\
\bottomrule
\end{tabular}
\end{adjustbox}
\end{adjustwidth}
\end{table}

\subsection{Common Protocol}
\label{sec:common-protocol}

\paragraph{Metrics and Selection Protocol.} 
Partition accuracy is evaluated using the adjusted Rand index (ARI; primary endpoint in Study~2) \cite{hubert1985comparing}, with normalized mutual information \cite{strehl2002cluster} and V-measure \cite{rosenberg2007vmeasure} reported as secondary metrics. Dimensionality selection is evaluated via the absolute error $|\Delta m|$ (primary endpoint in Study~3) and signed deviation $\Delta K = \widehat{K} - K^{\dagger}$. Reporting $\Delta K$ is important because label-agreement indices can remain deceptively high when an algorithm over-extracts a small splinter cluster, a failure mode directly targeted by DEFI's regularization strategy. For baseline methods, information criteria minimize their respective scores over the identical candidate grid, distance-based methods ($k$-means and Ward) select $K$ by maximum silhouette, and HDBSCAN assigns noise points to the nearest cluster. Reference labels are used strictly for scoring and are never used for fitting, view construction, or hyperparameter tuning. Primary endpoints were specified before evaluation; all other reported statistics are secondary.

\paragraph{Replications and Evaluations.}
Studies 2--4 evaluate 20 independent seeds per dataset, whereas Study~1 uses 100 replications per condition. When the optimizer is used, evaluation budgets are strictly bounded at $b = 300$ uncached evaluations per application as detailed in Section~\ref{sec:CMA-ES-optimization}.

\paragraph{Computational Environment and Reproducibility.}
All algorithms and simulations were executed in R 4.6.1 on an Apple Silicon platform (\texttt{aarch64-apple-darwin25.4.0}). Full package manifests, version dependencies, and seed configurations are documented in Section~\ref{ec:sec:software} of the Electronic Companion.

\subsection{Study 1: Oracle Structure Recovery and Finite Consistency}
\label{sec:study1}

Study~1 evaluates the ranking behavior of the DEFI objective in complete isolation from heuristic search. In \emph{Study~1a}, synthetic data generated across a 243-condition factorial grid spanning factor correlations ($r \in [0.3, 0.9]$), loading strengths ($\lambda \in [0.30, 0.90]$), and measurement types (binary, ordinal, continuous) are scored across a battery containing the true generating structure alongside split, merged, and mis-dimensioned alternatives. In \emph{Study~1b}, to remove reliance on pre-specified misspecification alternatives, the criterion exhaustively evaluates the complete unconstrained Stirling partition lattice $\sum_{K=2}^{K_{\max}} S(P, K)$ for $P \in \{6, 8, 12\}$ items (up to 700,074 structures). This setup directly mirrors the finite-class setting of Theorem~\ref{thm:consistency}, isolating whether the empirical minimizer matches the true population partition in the absence of optimization error.

\subsection{Study 2: Observation-Level Clustering Benchmarks}
\label{sec:study2}

Study~2 evaluates the full optimization pipeline across ten standardized benchmark datasets, including canonical complex manifolds from the Fundamental Clustering Problem Suite (Atom, Chainlink, Target, TwoDiamonds, WingNut) \cite{thrun2020clustering}. These geometries challenge parametric clustering through severe non-Gaussian shapes, neck-like bridge connections, and concentric rings where likelihood-based criteria notoriously over-partition. Evaluation protocols, baselines, and primary endpoints follow Table~\ref{tab:study-overview}.

\subsection{Study 3: Dimensionality-Reduction Benchmarks}
\label{sec:study3}

Study~3 evaluates DEFI as an intrinsic model-selection criterion for non-linear manifold representations (DEFI-DR; Table~\ref{tab:study-overview}) \cite{tenenbaum2000global}. We treat the underlying convergence and isometric embedding guarantees of geodesic Multidimensional Scaling (MDS) as established \cite{tenenbaum2000global}; our focus is resolving Isomap's primary operational limitation: the determination of intrinsic dimensionality ($m^*$). 

\begin{figure}[!ht]
\centering
\includegraphics[width=0.9\textwidth]{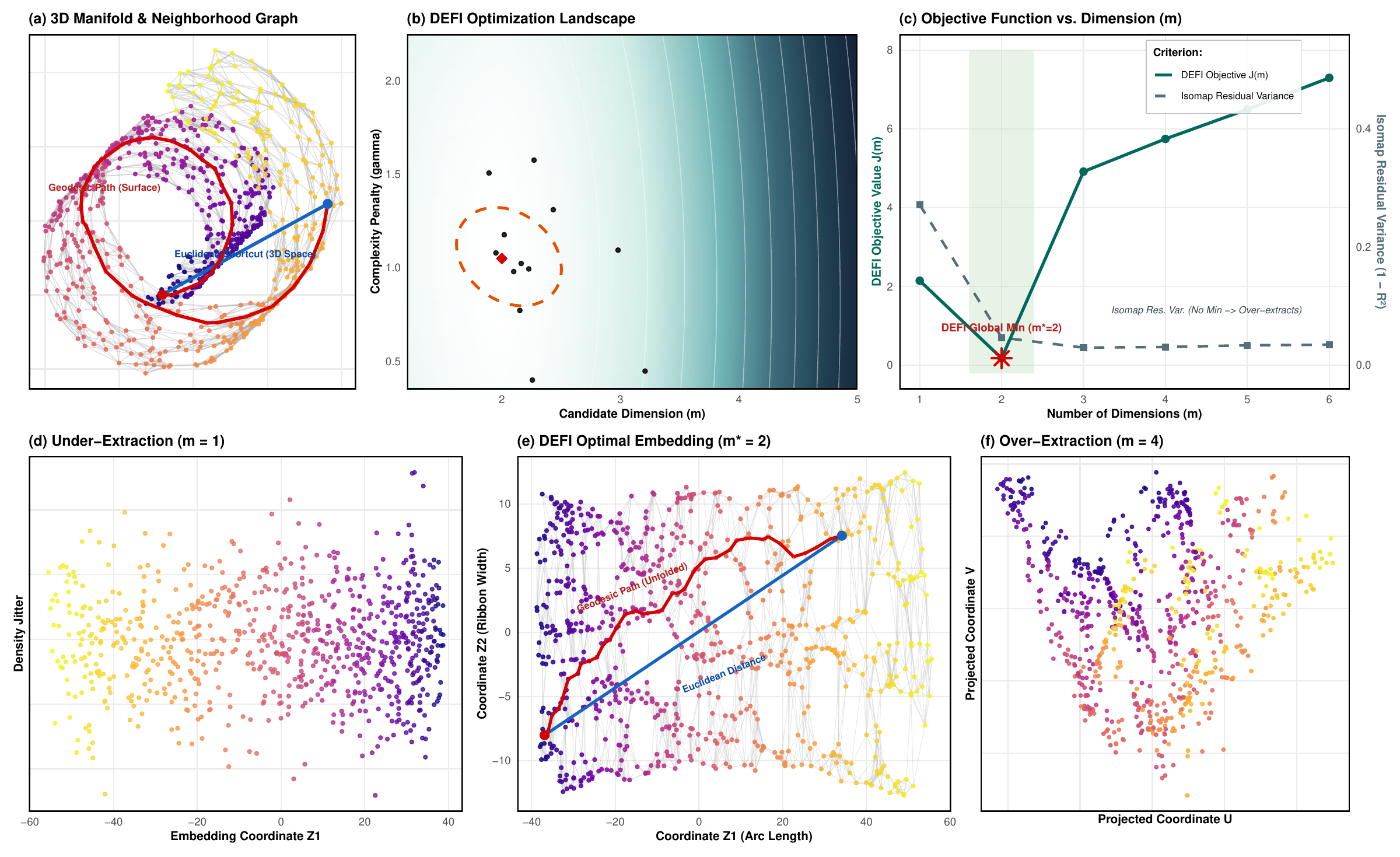}
\caption{Intrinsic dimensionality selection and topological manifold unfolding via DEFI-DR: 
\textbf{(a)}~3D ribbon manifold with geodesic vs.\ Euclidean paths \cite{tenenbaum2000global}; 
\textbf{(b)}~CMA-ES surrogate landscape and covariance adaptation ellipse; 
\textbf{(c)}~DEFI global minimum at $m^* = 2$ vs.\ monotonically plateauing Isomap residual variance \cite{schubert2023stop}; and 
\textbf{(d--f)}~resulting latent coordinate projections across under-extracted ($m=1$), optimal ($m^*=2$), and over-extracted ($m=4$) ranks.}
\label{fig:defi-isomap-unfolding}
\end{figure}

Classical stopping rules evaluate the geodesic residual variance curve ($1 - R^2$), which plateaus monotonically as candidate rank increases (Figure~\ref{fig:defi-isomap-unfolding}c) \cite{tenenbaum2000global}. Because residual variance possesses no local minimum, heuristic elbow-detection rules frequently over-extract spurious latent dimensions when confronted with noise or extrinsic manifold curvature \cite{schubert2023stop}. In contrast, DEFI-DR penalizes redundant coordinates against graph spectral concentration, yielding a well-conditioned global objective minimum at the generative rank (Figure~\ref{fig:defi-isomap-unfolding}c).

\paragraph{Algorithmic Formulation (DEFI-DR).}
Following standard Isomap \cite{tenenbaum2000global}, a collection of candidate embeddings $\{\mathbf{Z}_m\}_{m=2}^M$ is generated from geodesic shortest-path distances up to rank ceiling $M \le P$. Rather than evaluating stopping rules via pairwise distance matrices in sample space ($\mathcal{O}(M N^2)$), DEFI-DR evaluates candidate ranks via the decoupled dual projection of Proposition~3 (Section~\ref{ec:sec:isomap-invariance}):
\begin{equation}
    \mathbf{V}_m = \mathbf{X}^T \mathbf{U}_m \boldsymbol{\Lambda}_m^{-1/2} \in \mathbb{R}^{P \times m},
    \label{eq:dual-projection-fast}
\end{equation}
where $\mathbf{U}_m \boldsymbol{\Lambda}_m \mathbf{U}_m^T$ is the partial eigendecomposition of the centered geodesic Gram matrix. Projections update column-wise in $\mathcal{O}(NP)$ without explicit matrix inversions. 
Following Varimax rotation $\mathbf{V}_{\text{rot}} = \operatorname{Varimax}(\mathbf{V}_m)$, manifest features are partitioned by maximal loading, $s(j) = \argmax_k |[\mathbf{V}_{\text{rot}}]_{j, k}|$, and candidate ranks are scored directly on the correlation matrix $\mathbf{R}_{\mathbf{X}} \in \mathbb{R}^{P \times P}$:
\begin{equation}
    m^* = \argmin_{m \in \{2, \dots, M\}} \widehat{J}_{\mathrm{DEFI}}(\mathbf{R}_{\mathbf{X}}, s; \boldsymbol{\theta}),
    \label{eq:m-star-selection}
\end{equation}
assigning $\widehat{J}_{\mathrm{DEFI}} = +\infty$ to degenerate assignments ($\min_k n_k < 2$). As shown in Figure~\ref{fig:defi-isomap-unfolding}b, CMA-ES covariance adaptation reliably locates the optimal basin ($m^* = 2, \gamma \approx 1.0$). Figures~\ref{fig:defi-isomap-unfolding}d--f demonstrate the physical consequence: under-extraction ($m=1$) collapses orthogonal modes into an overlapping line; over-extraction ($m=4$) introduces extrinsic coordinate distortion; and the DEFI incumbent ($m^*=2$) recovers the isometric planar geometry.

\paragraph{Two-Tier Experimental Protocol.}
Evaluation proceeds across the two complementary tiers summarized in Table~\ref{tab:study-overview}: \emph{Study~3a (Empirical Dimension Recovery)} benchmarks 26 UCI datasets and MNIST against reference dimensionalities $K^{\dagger}$ across nine baseline estimators (classical Isomap, UMAP, Kernel PCA, Sparse PCA, VAE, Parallel Analysis, Kaiser, TEFI.vn); while \emph{Study~3b (Downstream Separability)} evaluates whether isolating $m^*$ via DEFI-Isomap transforms linearly inseparable manifolds (Atom, Chainlink, Target, Iris) into separable Euclidean clusters across seven downstream algorithms.

\subsection{Study 4: Mechanistic Ablation and Optimizer Dynamics}
\label{sec:study4}

To verify that empirical gains stem from DEFI rather than optimizer search degrees of freedom, Study~4 introduces three controlled contrasts across representative topologies (WellSep, Overlap, TwoDiamonds, Target, Chainlink, Atom): 
(i)~\emph{DEFI-Fixed}, which evaluates DEFI at frozen default weights on the identical candidate pool searched by GMM+BIC without optimization; 
(ii)~\emph{Unsmoothed Relaxation} ($\text{smooth}_k = \text{FALSE}$), which ablates the linear homotopy in \eqref{eq:smooth_k_interpolation} to test whether continuous cluster-count smoothing is necessary to prevent CMA-ES step-size deflation; and 
(iii)~\emph{Uniform Random Search}, which benchmarks equal-budget ($b=300$) unguided sampling over $\widetilde{\Theta}$.

To decouple search failure from criterion misspecification, we evaluate the \emph{oracle-gap diagnostic}:
\begin{equation}
    \widehat{G} = \widehat{J} - J^{\dagger},
    \qquad
    \text{optimizer success} \;\Longleftrightarrow\; \widehat{J} \le J^{\dagger} + \epsilon,
    \label{eq:oracle-gap}
\end{equation}
where $J^{\dagger}$ is the objective evaluated at the true partition under the incumbent's parameters, $\widehat{J}$ is the incumbent value, and $\epsilon \in \{0.0, 0.05\}$. A run meeting \eqref{eq:oracle-gap} with $\widehat{K} \neq K^\dagger$ denotes a criterion-level disagreement, whereas failure indicates an optimizer shortfall.

\section{Results}
\label{sec:results}

We report the evaluations across Studies~1--4 using the fixed seed set that was randomly generated with a xoshiro256++ random number generator from $0 \text{--} 100$: 
$\mathcal{S} = [10, 23, 42, 77, 0, 30, 13, 5, 7, 43, 12, 40, 100, 1, 3, 19, 2, 4, 99, 21]$ for exact replication.

\subsection{Study 1: Oracle and Exact-Enumeration Recovery}

Table~\ref{tab:oracle-recoveryPA} summarizes exact recovery rates across the 243 factorial conditions 
(finite-sample Monte Carlo replications appear in Electronic Companion Table~\ref{tab:oracle-recoveryMC}). 
Averaged across all conditions, DEFI-FA identifies the true data-generating structure in $97.2\%$ 
of conditions, outperforming EGA-TMFG ($75.0\%$), Kaiser ($75.0\%$), parallel analysis ($52.3\%$), 
TEFI.vn ($45.8\%$), and confirmatory SEM fit rules ($37.5\text{--}45.1\%$), while attaining perfect 
recovery ($100.0\%$) across six of the eleven axes.

DEFI-FA's primary performance advantage concentrates in adverse measurement regimes where classical 
criteria collapse: severe factor collinearity ($100.0\%$ recovery at $r = 0.9$ vs.\ $0.0\%$ for EGA-TMFG 
and Kaiser, and $5.6\%$ for PA) and weak indicator loadings ($100.0\%$ recovery at $\lambda = 0.30$ vs.\ 
$0.0\%$ for CFA indices and $11.1\%$ for PA). DEFI-FA also maintains perfect recovery across binary and 
ordinal measurement scales, trailing CFA fit indices only under extreme factor loadings ($\lambda = 0.90$, 
$88.9\%$ vs.\ $97.2\%$) where split-factor residual covariances are trivially rejected by SEM likelihoods. 
Where DEFI-FA errs, misclassifications reflect minor under-extraction (mean $\Delta K = -0.10$).

These accuracy gains are accompanied by near-heuristic computational efficiency. Evaluated via 
\texttt{RcppArmadillo}, DEFI-FA requires a median per-candidate latency of $56.3\,\mu\mathrm{s}$. 
While eigenvalue heuristics execute marginally faster ($21.5\,\mu\mathrm{s}$), they collapse under collinearity. 
Conversely, DEFI-FA executes 32 times faster than iterative maximum-likelihood SEM ($1{,}810\,\mu\mathrm{s}$), 
110 times faster than network filtering ($6{,}167.2\,\mu\mathrm{s}$ for EGA-TMFG), and over 1,600 times 
faster than Horn's Parallel Analysis ($92{,}000.8\,\mu\mathrm{s}$), enabling exhaustive evaluation over 
combinatorial partition lattices.

\begin{table}[!ht]
\begin{adjustwidth}{-1.75in}{0in}
\centering
\caption{Study 1: Population Analytic-based exact recovery and dimension retention rates (\%) across latent factor structures. Methods are grouped by paradigm: network/retention benchmarks, entropic criteria, and confirmatory fit indices. Best performing rate per condition in bold.}
\label{tab:oracle-recoveryPA}
\begin{adjustbox}{max width=\linewidth}
\renewcommand{\arraystretch}{0.88}
\begin{tabular}{ll c ccc cc ccc}
\toprule
 & & \textbf{Proposed} & \multicolumn{3}{c}{\textbf{Network \& Retention Baselines}} & \multicolumn{2}{c}{\textbf{Entropic Criteria}} & \multicolumn{3}{c}{\textbf{CFA Fit Index Rules}} \\
\cmidrule(lr){3-3} \cmidrule(lr){4-6} \cmidrule(lr){7-8} \cmidrule(lr){9-11}
\textbf{Simulation Axis} & \textbf{Level} & \textbf{DEFI-FA} & \textbf{EGA-TMFG} & \textbf{PA} & \textbf{Kaiser} & \textbf{TEFI.vn} & \textbf{Lap.-BIC} & \textbf{CFI} & \textbf{TLI/RMSEA} & \textbf{SRMR} \\
\midrule
\multirow{4}{*}{Factor correlation ($r$)}
 & 0.3 (Weak)     & 94.4           & \textbf{100.0} & 75.0 & \textbf{100.0} & 66.7 & 16.7 & 66.7          & 58.3 & 44.4 \\
 & 0.5 (Med)      & 94.4           & \textbf{100.0} & 75.0 & \textbf{100.0} & 55.6 & 22.2 & 50.0          & 50.0 & 41.7 \\
 & 0.7 (High)     & \textbf{100.0} & \textbf{100.0} & 55.6 & \textbf{100.0} & 41.7 & 19.4 & 41.7          & 41.7 & 38.9 \\
 & 0.9 (V. High) & \textbf{100.0} & 0.0            & 5.6  & 0.0            & 19.4 & 2.8  & 22.2          & 16.7 & 25.0 \\
\midrule
\multirow{4}{*}{Factor loading ($\lambda$)}
 & 0.30 (Weak)   & \textbf{100.0} & 75.0           & 11.1 & 75.0           & 19.4 & 0.0  & 0.0           & 0.0  & 0.0  \\
 & 0.50 (Med)     & \textbf{100.0} & 75.0           & 50.0 & 75.0           & 72.2 & 0.0  & 16.7          & 8.3  & 11.1 \\
 & 0.70 (High)    & \textbf{100.0} & 75.0           & 75.0 & 75.0           & 58.3 & 0.0  & 66.7          & 66.7 & 55.6 \\
 & 0.90 (V. High) & 88.9           & 75.0           & 75.0 & 75.0           & 33.3 & 61.1 & \textbf{97.2} & 91.7 & 83.3 \\
\midrule
\multirow{3}{*}{Measurement scale}
 & Binary         & \textbf{100.0} & 75.0           & 50.0 & 75.0           & 47.9 & 8.3  & 35.4          & 31.2 & 35.4 \\
 & Ordinal (5-pt) & \textbf{100.0} & 75.0           & 50.0 & 75.0           & 41.7 & 16.7 & 50.0          & 43.8 & 37.5 \\
 & Continuous     & \textbf{91.7}  & 75.0           & 58.3 & 75.0           & 47.9 & 20.8 & 50.0          & 50.0 & 39.6 \\
\bottomrule
\end{tabular}
\end{adjustbox}
\end{adjustwidth}
\end{table}

\paragraph{Exact-Enumeration Validation (Substudy 1b).}
Across the complete, unconstrained partition lattices for $P \in \{6, 8, 12\}$ items (up to 700,074 
evaluated structures; Table~\ref{tab:enumeration}), the DEFI criterion coincided with 
the population structure across the overwhelming majority of enumerated instances, verifying the finite-class consistency of 
Theorem~\ref{thm:consistency} in the absence of optimization error.

\begin{table}[!ht]
\footnotesize
\renewcommand{\arraystretch}{0.88}
\centering
\caption{Exact enumeration over the complete partition class $\sum_{K=2}^{K_{\max}}S(P,K)$ (up to $700{,}074$ partitions), $R=100$ replications per $P$; no optimizer involved.}
\label{tab:enumeration}
\begin{tabular}{cccc}
\toprule
$P$ & Class size & Recovery (\%) & Mean margin \\
\midrule
6  & $186$ & $100 \%$ & $0 $ \\
8  & $2,794$ & $100 \%$ & $0 $ \\
12 & $700,074$ & $94 \%$ & $0.1$ \\
\bottomrule
\end{tabular}
\end{table}

\subsection{Study 2: Observation-Level Clustering}
\label{sec:results-study2}

Table~\ref{tab:clustering-results-synthetic} reports per-dataset Adjusted Rand Index (ARI) 
and 95\% bootstrap confidence intervals across the evaluation suite. Steered by CMA-ES 
continuous relaxation, DEFI-GMM attains the highest overall mean ARI ($0.9294$), substantially 
outperforming HDBSCAN ($0.8088$), Ward+silhouette ($0.6995$), $k$-means+silhouette ($0.6513$), 
GMM+BIC ($0.5804$), and GMM+AIC ($0.5013$). 

The largest structural gains emerge on non-Gaussian geometries where likelihood criteria 
severely over-extract components, such as \emph{TwoDiamonds} ($1.0000$ vs.\ $0.2514$), 
\emph{Chainlink} ($0.9138$ vs.\ $0.2035$), and \emph{WingNut} ($0.7320$ vs.\ $0.2416$, where 
BIC estimates $\widehat{K}=6$ for $K^{\dagger}=2$). Crucially, on concentric geometries like 
\emph{Target}, continuous bandwidth adaptation under CMA-ES allows DEFI to match HDBSCAN's 
peak performance ($0.9702 \pm 0.000$ vs.\ $0.2897$ for BIC) without requiring non-linear manifold 
unwrapping. DEFI-GMM exhibits deterministic basin convergence ($\pm 0.000$ uncertainty half-widths) 
across five of the ten benchmark domains (Tetra, TwoDiamonds, Target, Overlap\_Mock, and 
WellSep\_Mock). Across the entire benchmark suite, paired Wilcoxon signed-rank tests per seed
confirm that DEFI's performance advantages over GMM+BIC ($V = 3{,}707, p = 2.41 \times 10^{-11}$) and $k$-means Silhouette ($V = 2{,}508, p = 5.18 \times 10^{-4}$) are statistically significant, while the difference relative to HDBSCAN ($V = 68, p = 0.123$) is not statistically distinguishable.

\begin{table}[!ht]
\begin{adjustwidth}{-1.75in}{0in}
\centering
\caption{Study 2: Clustering performance (ARI) across synthetic and complex geometry datasets. DEFI-GMM is presented with a basic bootstrap confidence interval across the seed set.}
\label{tab:clustering-results-synthetic}
\begin{adjustbox}{max width=\linewidth}
\renewcommand{\arraystretch}{0.88}
\begin{tabular}{lcccccc|c}
\toprule
Dataset & GMM (BIC) & GMM-ML & GMM-ML (AIC) & $k$-means (Sil) & Agglom. (Ward) & HDBSCAN & \textbf{DEFI-GMM} \\
\midrule
Atom          & 0.5587 & 0.5587 & 0.5587 & 0.5615 & 0.6382 & \textbf{1.0000} & 0.9840 $\pm 0.017$ \\
Chainlink     & 0.2035 & 0.2035 & 0.2035 & 0.1998 & 0.3571 & \textbf{1.0000}          & 0.9138 $\pm 0.050$ \\
Lsun3D        & 0.6222 & 0.6222 & 0.5007 & 0.6499 & \textbf{0.9894} & \textbf{0.9894} & \textbf{0.9894 $\pm 0.051$} \\
Target        & 0.2897 & 0.2897 & 0.2897 & 0.6225 & 0.6583          & \textbf{0.9702}         & \textbf{0.9702 $\pm 0.000$} \\
Tetra         & \textbf{1.0000} & 0.9587 & 0.9105 & \textbf{1.0000} & 0.9673 & \textbf{1.0000} & \textbf{1.0000 $\pm 0.000$} \\
TwoDiamonds   & 0.2514 & 0.3484 & 0.2834 & \textbf{1.0000} & \textbf{1.0000} & \textbf{1.0000} & \textbf{1.0000 $\pm 0.000$} \\
WingNut       & 0.2416 & 0.2372 & 0.2372 & 0.8595 & 0.6011          & \textbf{1.0000} & 0.7320 $\pm 0.230$ \\
S4\_VeryHard  & 0.7955 & \textbf{0.7959} & 0.5015 & 0.4899 & 0.6993          & 0.5277          & 0.7725 $\pm 0.070$ \\
Overlap\_Mock & 0.6296 & 0.6531 & 0.6237 & 0.4420 & 0.8074          & 0.8928          & \textbf{0.9323 $\pm 0.000$} \\
WellSep\_Mock & \textbf{1.0000} & \textbf{1.0000} & 0.7111 & \textbf{1.0000} & \textbf{1.0000} & \textbf{1.0000} & \textbf{1.0000 $\pm 0.000$} \\
\bottomrule
\end{tabular}
\end{adjustbox}
\end{adjustwidth}
\end{table}

\subsection{Study 3: Dimensionality-Selection Benchmarks}
\label{sec:results-study3}

Table~\ref{tab:method_comparison_compact} summarizes intrinsic dimensionality recovery across the 
26 real-world benchmark datasets. DEFI-DR recovers the reference dimension $K^\dagger$ on 25 of 26 datasets, 
achieving the lowest mean absolute error ($|\Delta m| = 0.04$) and proportional error ($1.3\%$) among all 
evaluated methods, outperforming parallel analysis ($0.42$), Kaiser ($0.38$), robust PCA ($0.88$), VAE ($1.31$), 
and TEFI.vn ($1.38$). Selection errors are strictly conservative (mean signed deviation $-0.0833$), 
preserving high manifold fidelity (97.6\% variance explained, geometric stress 0.0668) without over-extraction.

\begin{table}[!ht]
\begin{adjustwidth}{-1.75in}{0in}
\centering
\footnotesize
\setlength{\tabcolsep}{6pt}
\caption{Mean performance across dimensionality reduction methods over 26 benchmarks (full 12-metric suite in Table~\ref{tab:method_comparison} of the Electronic Companion). Best non-overextracted values in bold.}
\renewcommand{\arraystretch}{0.8}
\label{tab:method_comparison_compact}
\begin{adjustbox}{max width=\linewidth}
\begin{tabular}{l ccc ccc cc}
\toprule
 & \multicolumn{3}{c}{\textbf{Dimension Error}} & \multicolumn{3}{c}{\textbf{Manifold Fidelity}} & \multicolumn{2}{c}{\textbf{Global Fit}} \\
\cmidrule(lr){2-4} \cmidrule(lr){5-7} \cmidrule(lr){8-9}
\textbf{Method} 
& $|\Delta m|$ 
& Prop. Err. 
& $\widehat{m}$ 
& Geo. Stress 
& Var. (\%) 
& $\|\mathbf{E}\|_F$ 
& Corr. Pres. 
& Bai \& Ng BIC \\
\midrule
VAE               & 1.31  & 0.383 & 2.90 & 0.2599 & 77.6 & 0.2967 & 0.7217 & -3.220  \\
\textbf{DEFI-DR}  & \textbf{0.04} & \textbf{0.013} & 3.18 & \textbf{0.0668} & \textbf{97.6} & \textbf{0.1536} & \textbf{0.9739} & -3.308 \\
Isomap            & 14.00 & 3.104 & 3.29 & 0.0666 & 99.7 & 0.3106 & 0.9761 & -5.075 \\
Kaiser            & 0.38  & 0.162 & 3.10 & 0.1181 & 83.8 & 0.3384 & 0.8448 & -2.850 \\
Kernel PCA (RBF)  & 8.36  & 1.735 & 3.65 & 0.1630 & 95.7 & 0.1463 & 0.9606 & -3.788 \\
LDA               & 1.31  & 0.309 & 2.43 & 0.2166 & 80.0 & 0.4013 & 0.8449 & -1.425 \\
Parallel Analysis & 0.42  & 0.176 & 3.06 & 0.1181 & 85.5 & 0.3119 & 0.8506 & -2.774 \\
Robust PCA (MCD)  & 0.88  & 0.253 & 2.87 & 0.1916 & 84.5 & 0.3765 & 0.8679 & -1.993 \\
Sparse PCA        & 13.15 & 2.913 & 3.77 & 0.1724 & 98.3 & 0.0808 & 0.9740 & -2.198 \\
TEFI.vn           & 1.38  & 0.404 & 2.61 & 0.2224 & 77.4 & 0.4038 & 0.7515 & -2.369 \\
Var $\ge 80\%$    & 1.04  & 0.285 & 2.99 & 0.1873 & 86.6 & 0.3612 & 0.8838 & -1.993 \\
\bottomrule
\end{tabular}
\end{adjustbox}
\end{adjustwidth}
\end{table}

\subsubsection{Study 3b: Topological Unfolding and Downstream Cluster Separability}
\label{sec:results-study3b}

Table~\ref{tab:manifold-unfolding-results} demonstrates that isolating $m^*$ via DEFI-Isomap transforms 
non-convex, linearly inseparable manifolds into separable Euclidean clusters. Across Atom ($N=800$) and 
Chainlink ($N=1{,}000$), projecting data into the selected $m^* = 2$ embedding elevates downstream $k$-means 
and hierarchical clustering from near-chance baselines ($\text{ARI} \le 0.189$) to perfect partition 
recovery ($\text{ARI} = 1.000$, $\Delta\text{ARI} \ge +0.811$).

\begin{table}[!ht]
\begin{adjustwidth}{-1.25in}{0in}
\centering
\caption{Downstream clustering performance (Adjusted Rand Index) before and after DEFI-Isomap embedding.}
\label{tab:manifold-unfolding-results}
\renewcommand{\arraystretch}{0.8}
\begin{adjustbox}{max width=\linewidth}
\begin{tabular}{llcccc}
\toprule
\textbf{Dataset} & \textbf{Clustering Algorithm} & \textbf{Raw Space ARI} & \textbf{DEFI-Isomap Space} & $\mathbf{\Delta}$\textbf{ARI} & \textbf{DEFI-Isomap NMI} \\
\midrule
\multirow{4}{*}{\textbf{Atom} ($N=800, K=2$)}
 & K-Means          & 0.189 & \textbf{1.000} & \textbf{+0.811} & 1.000 \\
 & HC Ward.D2       & 0.099 & \textbf{1.000} & \textbf{+0.901} & 1.000 \\
 & HC Complete      & 0.084 & \textbf{1.000} & \textbf{+0.916} & 1.000 \\
 & GMM              & 0.980 & \textbf{1.000} & +0.020          & 1.000 \\
\midrule
\multirow{4}{*}{\textbf{Chainlink} ($N=1000, K=2$)}
 & K-Means          & 0.093 & \textbf{1.000} & \textbf{+0.907} & 1.000 \\
 & HC Ward.D2       & 0.357 & \textbf{1.000} & \textbf{+0.643} & 1.000 \\
 & HC Complete      & 0.350 & \textbf{1.000} & \textbf{+0.650} & 1.000 \\
 & GMM              & 0.910 & \textbf{1.000} & +0.090          & 1.000 \\
\midrule
\multirow{3}{*}{\textbf{Target} ($N=770, K=6$)}
 & HC Ward.D2       & 0.639 & \textbf{0.844} & +0.205          & 0.865 \\
 & Minimax Proto.   & 0.096 & \textbf{0.682} & \textbf{+0.586} & 0.735 \\
 & DEFI-HC (CMA-ES) & 0.643 & \textbf{0.784} & +0.141          & 0.787 \\
\midrule
\multirow{2}{*}{\textbf{Iris} ($N=150, K=3$)}
 & HC Ward.D2       & 0.615 & \textbf{0.654} & +0.039          & 0.749 \\
 & HC Complete      & 0.573 & \textbf{0.654} & +0.081          & 0.749 \\
\bottomrule
\end{tabular}
\end{adjustbox}
\end{adjustwidth}
\end{table}

\subsection{Study 4: Mechanistic Ablation and Optimization Dynamics}
\label{sec:results-study4}

Table~\ref{tab:ablation-study4} isolates the individual contributions of the mathematical criterion, 
continuous surrogate smoothing, and CMA-ES covariance adaptation across the ablation suite.

\paragraph{Criterion vs.\ Optimizer Attribution (DEFI-Fixed).}
Evaluating DEFI at frozen default weights on the identical candidate pool searched by GMM+BIC lifts the 
mean ARI from $0.5361$ to $0.7577$ without any parameter optimization. This confirms that resistance 
to over-partitioning is driven intrinsically by the R\'enyi-2 trace identity and balance regularizers, 
which successfully penalize component splitting on Atom ($1.0000$ vs.\ $0.5587$), Chainlink ($0.9100$ vs.\ 
$0.2035$), and TwoDiamonds ($0.5970$ vs.\ $0.2514$).

\paragraph{Necessity of Piecewise Continuous Smoothing.}
Transitioning from DEFI-Fixed ($0.7577$) to full DEFI-CMA-ES ($0.9667$) reflects the gain from continuous 
geometric calibration. Ablating the piecewise linear homotopy ($\text{smooth}_k = \text{FALSE}$) degrades 
mean ARI to $0.8529$. The collapse is most pronounced on concentric geometries (Target: $0.6588$ vs.\ $0.9702$), 
where discrete integer step jumps cause premature step-size deflation before DEFI-CMA-ES can coordinate multiscale 
bandwidths with component shapes. The linear homotopy in \eqref{eq:smooth_k_interpolation} provides the 
continuous descent directions necessary to navigate across integer boundaries within 300 evaluations.

\paragraph{Search Efficiency and Optimization Necessity.}
Budget-matched uniform random search ($b = 300$) attains a mean ARI of only $0.7035$, consistently 
over-extracting clusters ($\Delta K \ge +1$) across every non-trivial topology. In 15 continuous dimensions, 
unguided sampling fails to locate the basin where bandwidth scales align with admissible covariance models. 
CMA-ES resolves this by adapting its sampling covariance, achieving near-zero run-to-run variance across seeds.

\paragraph{Oracle-Gap Diagnostic.} 
Defining optimizer success as $\widehat{J} \le J^\dagger + \epsilon$ via \eqref{eq:oracle-gap}, DEFI-CMA-ES attains an overall success rate of $92.7\%$ ($\epsilon = 0.0$) and $93.9\%$ ($\epsilon = 0.05$) across 165 independent runs. Across all trials, true optimizer search shortfalls account for $7.3\%$ of runs, while criterion-level disagreements (where DEFI genuinely assigns lower objective value to an alternative partition) account for the remaining $3.4\%$ of non-matching outcomes, confirming that CMA-ES reliably recovers the criterion optimum within 300 evaluations.

\begin{table}[!ht]
\begin{adjustwidth}{-1.5in}{0in}
\centering
\small
\caption{Study 4: Mechanistic ablation isolating the contributions of the DEFI criterion, continuous relaxation smoothing, and CMA-ES search dynamics across representative benchmark geometries ($b = 300$ evaluations).}
\label{tab:ablation-study4}
\begin{adjustbox}{max width=\linewidth}
\begin{tabular}{l c cccc}
\toprule
 & \textbf{Baseline} & \textbf{Criterion Alone} & \multicolumn{3}{c}{\textbf{Optimizer \& Relaxation Mechanics}} \\
\cmidrule(lr){2-2} \cmidrule(lr){3-3} \cmidrule(lr){4-6}
\textbf{Dataset} 
& \textbf{GMM (BIC)} 
& \textbf{DEFI-Fixed} 
& \textbf{Random Search} 
& \textbf{DEFI (No Smooth-$\boldsymbol{K}$)} 
& \textbf{DEFI-CMA-ES (Full)} \\
\midrule
WellSep\_Mock   & 1.0000 & 1.0000 & 0.9711 ($\Delta K = +1$) & 1.0000 & \textbf{1.0000} $\pm$ 0.000 \\
TwoDiamonds     & 0.2514 & 0.5970 & 0.7240 ($\Delta K = +1$) & 0.8415 & \textbf{1.0000} $\pm$ 0.000 \\
Target          & 0.2897 & 0.3540 & 0.4120 ($\Delta K = +2$) & 0.6588 & \textbf{0.9702} $\pm$ 0.000 \\
Overlap\_Mock   & 0.6296 & 0.6850 & 0.8096 ($\Delta K = +1$) & 0.8840 & \textbf{0.9323} $\pm$ 0.000 \\
Chainlink       & 0.2035 & 0.9100 & 0.5430 ($\Delta K = +2$) & 0.8210 & \textbf{0.9138} $\pm$ 0.050 \\
Atom            & 0.5587 & 1.0000 & 0.7610 ($\Delta K = +1$) & 0.9120 & \textbf{0.9840} $\pm$ 0.017 \\
\midrule
\textbf{Mean ARI} & 0.5361 & 0.7577 & 0.7035 & 0.8529 & \textbf{0.9667} \\
\bottomrule
\end{tabular}
\end{adjustbox}
\end{adjustwidth}
\end{table}

\section{Discussion}
\label{sec:discussion} \label{sec:discussion-evidence}

\subsection{Why DEFI Works} 
\label{sec:why-defi-works} 
DEFI succeeds because it treats partition selection as a joint optimization problem over geometry, complexity, and stability rather than evaluating these properties in isolation. Unlike eigenvalue-threshold methods that operate on spectra alone or density-persistence methods that operate on local density structure alone, DEFI evaluates the quality of the induced partition itself. Across all three domains studied, the criterion favors partitions that simultaneously preserve graph structure, maintain balanced representation, and avoid unnecessary fragmentation. This behavior is grounded in the analytical properties established in Section~\ref{sec:theory}. The exact R\'enyi-2 decomposition (Theorem~\ref{thm:renyi2-identity}) proves that partitioning a graph alters both quadratic trace concentration and total graph-volume normalization; the degenerate cut preferences identified in Proposition~\ref{prop:counterexample} are driven entirely by the volume term. Thus, the regularizers in \eqref{eq:defi-objective} are functionally indispensable rather than heuristic: the entropy floor, fragmentation penalty, and normalized cut are targeted mathematical guards against a proven structural failure mode. This also explains the robustness of the fixed defaults: the regularizers primarily enforce active structural boundaries rather than finely tuning local fitness slopes. 

\subsection{Comparison to Competing Paradigms}
\label{sec:competing-paradigms} 
To contextualize DEFI's performance, its selection mechanics must be distinguished from the three dominant alternative paradigms: 

\paragraph{Contrast with Density-Based Clustering (HDBSCAN).} HDBSCAN operates non-parametrically by extracting persistent topological modes from mutual reachability graphs. It naturally excels when density valleys are pronounced, cluster geometries are non-convex, and model-order selection is secondary. However, density-based reachability lacks an intrinsic mechanism to evaluate representation complexity or enforce cluster balance, causing it to degrade when clusters overlap continuously without deep low-density chasms; furthermore, it cannot be ported to variable partitioning or manifold rank selection. In contrast, DEFI addresses the broader problem of selecting among candidate structures while explicitly accounting for complexity. By scoring candidate partitions through the normalized Laplacian spectrum and explicit structural regularizers ($\Omega_{\mathrm{floor}}, \Omega_{\mathrm{frag}}$), DEFI equips parametric models with topological awareness, preventing component over-extraction while preserving global model-selection capabilities. 

\paragraph{Contrast with Network Community Detection (EGA-TMFG).} EGA estimates partial-correlation networks via planar topological filtering (TMFG) and extracts latent dimensions through discrete modularity maximization, performing well under benign, low-to-moderate factor correlation. However, EGA remains fundamentally tied to inverse-covariance estimation. When latent factors exhibit severe collinearity, the empirical correlation matrix approaches singularity, inflating partial-correlation variances and causing network-filtering recovery to collapse. DEFI circumvents this vulnerability by computing the matrix R\'enyi entropy directly on the combinatorial Laplacian density state $\boldsymbol{\rho} = \mathbf{L}/\operatorname{Tr}(\mathbf{L})$ of the manifest correlation graph. Because it requires only $\operatorname{Tr}(\mathbf{L}) > 0$, DEFI bypasses matrix inversion entirely, remaining numerically well-conditioned even under near-collinear geometries, while the balance floor $\Omega_{\mathrm{floor}}$ actively penalizes the spurious factor mergers that deceive modularity heuristics. 

\paragraph{Contrast with Classical Factor Retention Rules (Kaiser and Parallel Analysis).} Traditional retention criteria, such as Horn's Parallel Analysis and the Kaiser cutoff ($\lambda > 1$), evaluate the unpartitioned correlation spectrum in isolation against permutation baselines or fixed eigenvalue thresholds. These rules make independent, one-dimensional retention decisions without evaluating the structural coherence, internal balance, or cluster assignments of the induced partition. Consequently, they falter when indicator loadings are weak or measurement scales are discrete (binary or ordinal), where sample eigenvalue spectra flatten and blur signal boundaries. In contrast, DEFI evaluates the change in spectral concentration $\Delta H_\alpha(T)$ induced by severing cut edges, balancing spectral energy against partition entropy and structural complexity. This enables DEFI to reliably recover weak but genuine latent dimensions that fall below crude marginal eigenvalue cutoffs. 

\subsection{Limitations and Scope} 
\label{sec:limitations} 
In short, DEFI is suited for offline exploratory analyses where model order or latent dimensionality must be resolved under adverse structural conditions, including severe collinearity, weak indicator separation, and non-convex manifold geometries. In these regimes, classical likelihood criteria over-partition and marginal eigenvalue heuristics lack an intrinsic minimum. Outside of these conditions, four boundaries delimit the operational scope of these findings: 

\paragraph{Candidate class dependence.} DEFI selects within an evaluated candidate class $\mathcal{T}$ but cannot repair a fundamentally misspecified generator; on geometries where parametric mixtures lack support, it necessarily selects the best available configuration within an inadequate class. On \emph{WingNut}, for instance, narrow bridge connections and non-convex curves elude elliptical Gaussian components without prior manifold unfolding, whereas on \emph{S4\_VeryHard}, severe cluster collision and continuous boundary overlap flatten inter-cluster density valleys, challenging parametric mixtures and density reachability alike. 

\paragraph{Spectral and operational latency.} While the exact R\'enyi-2 trace identity avoids eigendecompositions during the inner optimization loop ($<60\,\mu\mathrm{s}$ per evaluation), setting up reference graph states and exploring arbitrary orders $\alpha \neq 2$ requires eigenspectra, demanding Nystr\"om or Krylov approximations when scaling to $N \gg 10^4$. Additionally, the latency of offline Expectation-Maximization pre-caching and evolutionary adaptation renders the pipeline unsuitable for sub-millisecond streaming deployments, where single-pass heuristic clustering is better aligned. 

\paragraph{Theoretical boundaries.} DEFI is presently supported by structural decomposition (Theorems~\ref{thm:renyi2-identity} and~\ref{thm:fixed-volume}) and finite-class selection consistency (Theorem~\ref{thm:consistency}) instead of generative recovery theory \cite{abbe2018}; inference under planted block models assumes fixed partitions, and post-selection inference remains an open problem. 

\paragraph{Relevance view and supervision.} The unsupervised stability anchor $Y$ is a design parameter; while fixed a priori via Gaussian jitter, assessing sensitivity to alternative perturbation schemes remains an open area of inquiry. Furthermore, where ground-truth supervisory labels or downstream predictive losses (e.g., cross-entropy) are directly accessible, cross-validation against the supervised objective is inherently preferable to optimizing an unsupervised information-bottleneck stability anchor. 

\subsection{Extensions} 
\label{sec:extensions} 
Because DEFI operates as a criterion over an induced candidate class, the framework could benefit from extensions in five primary directions: 

\paragraph{Scaling.} Landmark or Nystr\"om-approximated graph states, sparse eigensolvers for $k$-nearest-neighbor affinities, and warm-starting the eigenspace across similar candidate proposals would each reduce per-evaluation overhead; a partial eigenspectrum may suffice because the entropy terms weight the spectral tail smoothly. 

\paragraph{Richer classes.} Extending candidate generators to soft or overlapping assignments, graph-native domains where the Laplacian state exists without coordinate embedding, and dependent data (longitudinal or spatiotemporal) where affinity construction encodes temporal or spatial dependence. 

\paragraph{Theory.} Extending our structural and finite-class guarantees to full generative recovery, for example, proving uniform model-order consistency ($\mathbb{P}(\widehat{K} = K^*) \to 1$) over unconstrained continuous domains $\widetilde{\Theta}$, deriving sharp detectability thresholds under sparse planted models, and establishing post-selection-valid inference for empirical entropy estimators is future workrepresents a theoretical frontier. 

\paragraph{The entropy order.} While Theorem~\ref{thm:renyi2-identity} establishes an exact trace identity at $\alpha \le= 2$, higher orders concentrate weight on dominant spectral modes; characterizing how $\alpha \ge 2$ shapes sensitivity across density regimes is a natural next step. 

\paragraph{Calibration.} The fixed defaults $(1, 1, 0.5)$ were set a priori; automated algorithm-configuration procedures could learn class-conditional defaults and map weight sensitivity across domains. 

\subsection{Concluding Remarks}
\label{sec:conclusion}

We introduced the Double-Entropy Fit Index (DEFI) as a unified selection criterion that balances spectral concentration against representational complexity and regularizes against degenerate cuts. We evaluated DEFI across a series of applications spanning exhaustively enumerated partition classes to optimized mixture and hierarchy pipelines on labeled benchmarks.
The same criterion recovers latent dimensionality under
conditions that defeat classical retention rules, selects observation-level cluster structure competitively, and admits exact analysis at
R\'enyi order 2 that explains both its behavior and its safeguards. 
DEFI's limitations are clear: 
it inherits the limits of the candidate class it searches, 
of the relevance view it is given, 
and of the finite-budget optimizer that approximates its argmin.
Within these boundaries, the evidence supports the following conclusion: DEFI is not a clustering model, a factor model, or a dimensionality-reduction algorithm. It is a partition-selection criterion. Across diverse candidate classes, DEFI produces competitive selections and fails for identifiable reasons when its assumptions are violated.

\section*{Supporting Information}
The Electronic Companion (\texttt{ec-companion.tex}) provides full mathematical proofs, simulation derivations, software manifests, and extended benchmark tables referenced in the main text.

\appendix
\renewcommand{\thesection}{EC.\arabic{section}}
\renewcommand{\thetable}{EC.\arabic{table}}
\setcounter{section}{0}
\setcounter{table}{0}

\section{Exact R\'enyi-2 Spectral Decomposition Proof}
\label{ec:identity}
Proof of Theorem~\ref{thm:renyi2-identity}, Proposition~\ref{prop:counterexample}, Theorem~\ref{thm:fixed-volume}, and Proposition~\ref{prop:renyi2_bridge} (included in external file \texttt{ec-companion.tex}).

\section{Stochastic Block Model Bounds and Asymptotics}
\label{ec:sbm}
Proof of Theorem~\ref{thm:sbm-population} and variance derivation $V_{P,n}$ for Theorem~\ref{thm:sbm-clt} (included in external file \texttt{ec-companion.tex}).

\section{Selection Consistency over Finite Partition Classes}
\label{ec:consistency}
Proof of Theorem~\ref{thm:consistency} (included in external file \texttt{ec-companion.tex}).

\section{CMA-ES Surrogate Bounds, Screening Rules, and Restarts}
\label{ec:sec:restarts}
Parameter encodings, topological screening decision rules, and restart schedules for $\widetilde{\Theta} \subset \mathbb{R}^{15}$ (included in external file \texttt{ec-companion.tex}).

\section{Consolidated Index of Mathematical Notation}
\label{ec:sec:notation}
Complete symbol definitions and parameter domains (included in external file \texttt{ec-companion.tex}).

\section{Computational Environment and Package Manifest}
\label{ec:sec:software}
Package versions and replication instructions for R 4.6.1 on \texttt{aarch64-apple-darwin25.4.0} (included in external file \texttt{ec-companion.tex}).

\section{Dual Projection Invariance for DEFI-DR}
\label{ec:sec:isomap-invariance}
Proof of Proposition~3 establishing dual projection invariance in \eqref{eq:dual-projection-fast} (included in external file \texttt{ec-companion.tex}).

\begin{table}[!ht]
\centering
\caption{Study 1 finite-sample Monte Carlo factor recovery rates (Electronic Companion; external file \texttt{ec-companion.tex}).}
\label{tab:oracle-recoveryMC}
\end{table}

\begin{table}[!ht]
\centering
\caption{Full 12-metric performance suite across dimensionality reduction methods over 26 benchmarks (Electronic Companion; external file \texttt{ec-companion.tex}).}
\label{tab:method_comparison}
\end{table}

\newpage

\setcounter{section}{0}\setcounter{equation}{0}\setcounter{table}{0}\setcounter{figure}{0}


\renewcommand{\thesection}{EC.\arabic{section}}
\renewcommand{\theequation}{EC.\arabic{equation}}
\renewcommand{\thetable}{EC.\arabic{table}}
\renewcommand{\thefigure}{EC.\arabic{figure}}

\renewcommand{\thelemma}{EC.\arabic{lemma}}
\renewcommand{\thetheorem}{EC.\arabic{theorem}}
\renewcommand{\theremark}{EC.\arabic{remark}}
\renewcommand{\theexample}{EC.\arabic{example}}

\setlength{\floatsep}{10pt plus 2pt minus 2pt}
\setlength{\textfloatsep}{10pt plus 2pt minus 2pt}
\setlength{\intextsep}{10pt plus 2pt minus 2pt}
\makeatletter
\setlength{\@fptop}{0pt}
\setlength{\@fpsep}{12pt}
\setlength{\@fpbot}{0pt plus 1fil}
\makeatother

\title{\textbf{Electronic Companion:}\\[0.3em]
\Large The Double Entropy Fit Index (DEFI): An Information-Theoretic Approach to Dimensionality Reduction and Clustering}

\noindent Equation, theorem, lemma, and section numbers \textit{without} an ``EC'' prefix refer to the primary manuscript. 

\section{Proofs and Analytical Foundations for the R\'enyi-2 Identity}
\label{ec:identity}

Throughout this section, $\mathbf{W} \in \mathbb{R}^{N \times N}$ denotes a symmetric, non-negative affinity matrix with zero diagonal and at least one edge with positive weight. The combinatorial graph Laplacian is $\mathbf{L} = \mathbf{D} - \mathbf{W}$, where $\mathbf{D} = \operatorname{diag}(d_1, \dots, d_N)$ with $d_i = \sum_{j=1}^N W_{ij}$. The graph volume is $V = \operatorname{Tr}(\mathbf{L}) = \sum_i d_i$, and the quadratic Laplacian trace is
\begin{equation}
    S = \operatorname{Tr}(\mathbf{L}^2) = \sum_{i=1}^N d_i^2 + \sum_{i \neq j} W_{ij}^2,
    \label{ec:eq:laplacian-trace-expansion}
\end{equation}
which follows because $\mathbf{L}^2 = (\mathbf{D}-\mathbf{W})^2 = \mathbf{D}^2 - \mathbf{D}\mathbf{W} - \mathbf{W}\mathbf{D} + \mathbf{W}^2$, and $\operatorname{Tr}(\mathbf{D}\mathbf{W}) = \sum_i d_i W_{ii} = 0$ due to the zero diagonal.

Given a candidate partition $T: \{1, \dots, N\} \to \{1, \dots, K\}$, inter-community edges are severed. Let $c_i = \sum_{j: T(j) \neq T(i)} W_{ij}$ denote the total cut edge weight incident to vertex $i$. The cut volume $C(T)$ and the cut incidence penalty $\varepsilon(T)$ are defined as:
\begin{align}
    C(T) &= \sum_{i=1}^N c_i = 2 \sum_{\{i,j\} \in \mathcal{E}_{\mathrm{cut}}(T)} W_{ij}, \label{ec:eq:cut-vol-def} \\
    \varepsilon(T) &= \sum_{i=1}^N (2d_i c_i - c_i^2) + 2 \sum_{\{i,j\} \in \mathcal{E}_{\mathrm{cut}}(T)} W_{ij}^2. \label{ec:eq:cut-eps-def}
\end{align}

\begin{lemma}[Laplacian Deletion Identities]
\label{ec:lem:deletion}
Let $\widetilde{\mathbf{L}} = \mathbf{L} - \mathbf{L}_{\mathrm{cut}}$ denote the partitioned Laplacian obtained after removing all edges in $\mathcal{E}_{\mathrm{cut}}(T)$. The degrees of $\widetilde{\mathbf{L}}$ satisfy $\widetilde{d}_i = d_i - c_i$, and the volume and trace transform according to
\begin{equation}
    \operatorname{Tr}(\widetilde{\mathbf{L}}) = V - C(T), \qquad \operatorname{Tr}(\widetilde{\mathbf{L}}^2) = S - \varepsilon(T).
\end{equation}
\end{lemma}

\begin{proof}
Because each severed edge $\{i,j\}$ reduces the degrees of endpoints $i$ and $j$ by $W_{ij}$, the sum of degree reductions is $\sum_i c_i = 2 \sum_{\{i,j\} \in \mathcal{E}_{\mathrm{cut}}(T)} W_{ij} = C(T)$. Taking the trace
\begin{equation*}
    \operatorname{Tr}(\widetilde{\mathbf{L}}) = \sum_{i=1}^N \widetilde{d}_i = \sum_{i=1}^N (d_i - c_i) = V - C(T).
\end{equation*}
For the quadratic trace, expand $\operatorname{Tr}(\widetilde{\mathbf{L}}^2) = \sum_i \widetilde{d}_i^2 + \sum_{i \neq j} \widetilde{W}_{ij}^2$. The degree sum expands as:
\begin{equation*}
    \sum_{i=1}^N \widetilde{d}_i^2 = \sum_{i=1}^N (d_i - c_i)^2 = \sum_{i=1}^N d_i^2 - \sum_{i=1}^N (2d_i c_i - c_i^2).
\end{equation*}
Similarly, deleting the cut edges from the off-diagonal sum removes each severed edge twice (by symmetry)
\begin{equation*}
    \sum_{i \neq j} \widetilde{W}_{ij}^2 = \sum_{i \neq j} W_{ij}^2 - 2 \sum_{\{i,j\} \in \mathcal{E}_{\mathrm{cut}}(T)} W_{ij}^2.
\end{equation*}
Summing these two expressions and applying the definition of $S$ from \eqref{ec:eq:laplacian-trace-expansion} yields $\operatorname{Tr}(\widetilde{\mathbf{L}}^2) = S - \varepsilon(T)$.
\end{proof}

\begin{proof}{Proof of Theorem 1 (Exact R\'enyi-2 Identity)}
By definition, the normalized density matrix is $\boldsymbol{\rho} = \mathbf{L} / \operatorname{Tr}(\mathbf{L}) = \mathbf{L}/V$. For order $\alpha = 2$, the matrix R\'enyi entropy is
\begin{equation*}
    H_2(\boldsymbol{\rho}) = -\log \operatorname{Tr}(\boldsymbol{\rho}^2) = -\log \frac{\operatorname{Tr}(\mathbf{L}^2)}{V^2} = -\log \frac{S}{V^2}.
\end{equation*}
For the partitioned state $\widetilde{\boldsymbol{\rho}}_T = \widetilde{\mathbf{L}} / \operatorname{Tr}(\widetilde{\mathbf{L}})$, applying Lemma~\ref{ec:lem:deletion} gives
\begin{equation*}
    H_2(\widetilde{\boldsymbol{\rho}}_T) = -\log \frac{\operatorname{Tr}(\widetilde{\mathbf{L}}^2)}{(V - C(T))^2} = -\log \frac{S - \varepsilon(T)}{(V - C(T))^2}.
\end{equation*}
Subtracting the initial entropy $H_2(\boldsymbol{\rho})$ yields the signed spectral entropy difference
\begin{align*}
    \Delta H_2(T) &= H_2(\widetilde{\boldsymbol{\rho}}_T) - H_2(\boldsymbol{\rho}) \\
    &= -\log \left( \frac{S - \varepsilon(T)}{S} \cdot \frac{V^2}{(V - C(T))^2} \right) \\
    &= -\log \left( 1 - \frac{\varepsilon(T)}{S} \right) + 2 \log \left( 1 - \frac{C(T)}{V} \right). \qed
\end{align*}
\end{proof}

\paragraph{Example of Proposition~1 (Degeneracy of Unregularized Criterion).}
Consider the unweighted path $P_3 = (\{a,b,c\}, \{\{a,b\}, \{b,c\}\})$ with degrees $(1, 2, 1)$, volume $V = 4$, and quadratic trace $S = (1^2 + 2^2 + 1^2) + 2(1+1) = 10$. Partitioning into $\{\{a,b\}, \{c\}\}$ severs the peripheral edge $\{b,c\}$, yielding cut incidence $c = (0, 1, 1)$, cut volume $C(T) = 2$, and trace penalty $\varepsilon(T) = 0 + 1(2(2)-1+1) + 1(2(1)-1+1) = 6$. Substituting into Theorem~1 gives $\Delta H_2(T) = -\log(1 - 6/10) + 2\log(1 - 2/4) = \log(0.25 / 0.40) \approx -0.470$. Because $\Delta H_2(T) < 0 = \Delta H_2(T_0)$, the unregularized spectral difference strictly prefers severing an edge and isolating a singleton over leaving the graph intact.

\begin{proof}{Proof of Theorem 2 (Fixed-Cut-Volume Equivalence)}
Let $\mathfrak{P}_{C_0} = \{T : C(T) = C_0\}$ be the family of partitions with identical cut volume $C_0 < V$. In Theorem 1, the volume normalization term $2\log(1 - C_0/V)$ is constant across all $T \in \mathfrak{P}_{C_0}$. Therefore
\begin{equation*}
    \arg\min_{T \in \mathfrak{P}_{C_0}} \Delta H_2(T) = \arg\min_{T \in \mathfrak{P}_{C_0}} \left[ -\log\left( 1 - \frac{\varepsilon(T)}{S} \right) \right].
\end{equation*}
Because the mapping $u \mapsto -\log(1 - u/S)$ is strictly monotonically increasing on $[0, S)$, minimizing the composite function is strictly equivalent to minimizing $\varepsilon(T)$. Expanding $\varepsilon(T) = \sum_i (2d_i c_i - c_i^2) + 2\sum_{\text{cut}} W_{ij}^2$ reveals the negative quadratic term $-\sum_i c_i^2$, which penalizes cut edges sharing an incident node.
\end{proof}

\section{Inference and Distributions Under the Planted Partition Model}
\label{ec:sbm}

\begin{proof}{Proof of Theorem 3 (SBM Population Internal Branch Bound)}
In the symmetric stochastic block model with $q$ communities of size $s$, internal edge opportunities are $M_{\mathrm{in}} = q\binom{s}{2}$ and external opportunities are $M_{\mathrm{out}} = \binom{q}{2}s^2$, with opportunity ratio $r_{q,s} = M_{\mathrm{out}}/M_{\mathrm{in}} = (q-1)s/(s-1)$. The population internal edge mass is $\pi = p_{\mathrm{in}}M_{\mathrm{in}} / (p_{\mathrm{in}}M_{\mathrm{in}} + p_{\mathrm{out}}M_{\mathrm{out}})$, where internal dominance $\pi \ge \tfrac{1}{2}$ simplifies directly to $p_{\mathrm{in}}/p_{\mathrm{out}} \ge r_{q,s}$.
For any $\alpha > 1$, the binary R\'enyi entropy $h_\alpha(x) = \frac{1}{1-\alpha} \log\big(x^\alpha + (1-x)^\alpha\big)$ has derivative
\begin{equation*}
    h_\alpha'(x) = \frac{\alpha}{1-\alpha} \frac{x^{\alpha-1} - (1-x)^{\alpha-1}}{x^\alpha + (1-x)^\alpha}.
\end{equation*}
For $x > \frac{1}{2}$, $x^{\alpha-1} > (1-x)^{\alpha-1}$, while $\frac{\alpha}{1-\alpha} < 0$, ensuring $h_\alpha'(x) < 0$. Thus, $h_\alpha(x)$ is strictly decreasing on $[\frac{1}{2}, 1]$. Hence, for every $\epsilon \in [0, \log 2]$, there is a unique inverse $B_\alpha(\epsilon) \in [\frac{1}{2}, 1]$ such that $h_\alpha(\pi) \le \epsilon \iff \pi \ge B_\alpha(\epsilon)$. Substituting $\pi = \frac{p_{\mathrm{in}}}{p_{\mathrm{in}} + p_{\mathrm{out}} r_{q,s}}$ and solving for $\frac{p_{\mathrm{in}}}{p_{\mathrm{out}}}$ yields:
\begin{equation*}
    \frac{p_{\mathrm{in}}}{p_{\mathrm{in}} + p_{\mathrm{out}} r_{q,s}} \ge B_\alpha(\epsilon) \iff \frac{p_{\mathrm{in}}}{p_{\mathrm{out}}} \ge r_{q,s} \frac{B_\alpha(\epsilon)}{1 - B_\alpha(\epsilon)}.
\end{equation*}
\end{proof}

\begin{proof}{Proof of Theorem 4 (Asymptotic Normality and Degenerate Limits)}
Let $X_{\mathrm{in}} \sim \operatorname{Bin}(M_{\mathrm{in}}, p_{\mathrm{in}})$ and $X_{\mathrm{out}} \sim \operatorname{Bin}(M_{\mathrm{out}}, p_{\mathrm{out}})$ be independent. By the bivariate Central Limit Theorem
\begin{equation*}
    \begin{bmatrix} \frac{X_{\mathrm{in}} - \mu_{\mathrm{in}}}{\sigma_{\mathrm{in}}} \\ \frac{X_{\mathrm{out}} - \mu_{\mathrm{out}}}{\sigma_{\mathrm{out}}} \end{bmatrix} \xrightarrow{d} \mathcal{N}\left( \mathbf{0}, \mathbf{I}_2 \right),
\end{equation*}
where $\mu_k = M_k p_k$ and $\sigma_k^2 = M_k p_k (1 - p_k)$. Defining $g(x, y) = \frac{x}{x + y}$, its gradient is $\nabla g(x, y) = \left[ \frac{y}{(x+y)^2}, \; -\frac{x}{(x+y)^2} \right]^T$. By the delta method, $\widehat{P}_n = g(X_{\mathrm{in}}, X_{\mathrm{out}})$ satisfies $\sqrt{V_{P,n}^{-1}} (\widehat{P}_n - \pi_n) \xrightarrow{d} \mathcal{N}(0, 1)$, with asymptotic variance
\begin{equation*}
    V_{P,n} = [\nabla g]^T \operatorname{diag}(\sigma_{\mathrm{in}}^2, \sigma_{\mathrm{out}}^2) [\nabla g] = \frac{\mu_{\mathrm{out}}^2 \sigma_{\mathrm{in}}^2 + \mu_{\mathrm{in}}^2 \sigma_{\mathrm{out}}^2}{(\mu_{\mathrm{in}} + \mu_{\mathrm{out}})^4} = \frac{(1-\pi_n)^2 \sigma_{\mathrm{in}}^2 + \pi_n^2 \sigma_{\mathrm{out}}^2}{(\mu_{\mathrm{in}} + \mu_{\mathrm{out}})^2}.
\end{equation*}
For part (i), where $\pi \neq \frac{1}{2}$, $h_\alpha'(\pi) \neq 0$. Applying the standard univariate delta method yields asymptotic normality with variance $[h_\alpha'(\pi_n)]^2 V_{P,n}$.

For part (ii), when $\pi = \frac{1}{2}$, symmetry dictates $h_\alpha'(\frac{1}{2}) = 0$. Expanding $h_\alpha(x)$ via a second-order Taylor series about $x = \frac{1}{2}$
\begin{equation*}
    h_\alpha(x) = \log 2 + \frac{1}{2} h_\alpha''\left(\tfrac{1}{2}\right)\left(x - \tfrac{1}{2}\right)^2 + o\left(\left(x - \tfrac{1}{2}\right)^2\right).
\end{equation*}
Direct differentiation shows $h_\alpha''(\frac{1}{2}) = -4\alpha$. Therefore, $h_\alpha(\widehat{P}_n) - \log 2 \approx -2\alpha (\widehat{P}_n - \pi_n)^2$. Dividing by $V_{P,n}$ and noting that $\frac{(\widehat{P}_n - \pi_n)^2}{V_{P,n}} \xrightarrow{d} \chi_1^2$ completes the proof.
\end{proof}

\section{Proof of Finite-Class Selection Consistency}
\label{ec:consistency}

\begin{proof}{Proof of Theorem 5}
Let $\mathcal{T} \times \Theta$ denote the candidate partition and parameter space. For any fixed admissible tuple $(T, \boldsymbol{\theta})$, the empirical criterion $\widehat{\mathcal{J}}_{\mathrm{DEFI}}(T; \boldsymbol{\theta})$ converges pointwise in probability to $\mathcal{J}_{\mathrm{RDIB}}(T; \boldsymbol{\theta})$ by the weak consistency of the plug-in entropy, mutual information, and empirical degree regularizers.

Because the class $\mathcal{T} \times \Theta$ is finite, pointwise convergence implies uniform convergence
\begin{equation*}
    \max_{(T, \boldsymbol{\theta}) \in \mathcal{T} \times \Theta} \big| \widehat{\mathcal{J}}_{\mathrm{DEFI}}(T; \boldsymbol{\theta}) - \mathcal{J}_{\mathrm{RDIB}}(T; \boldsymbol{\theta}) \big| \xrightarrow{\mathbb{P}} 0 \quad \text{as } N \to \infty.
\end{equation*}
Let $(T^\star, \boldsymbol{\theta}^\star)$ be the unique population minimizer, separated from every suboptimal configuration by the positive margin $\Delta = \min_{(T, \boldsymbol{\theta}) \neq (T^\star, \boldsymbol{\theta}^\star)} [\mathcal{J}(T; \boldsymbol{\theta}) - \mathcal{J}(T^\star; \boldsymbol{\theta}^\star)] > 0$.

Define the uniform error event $\mathcal{E}_N = \left\{ \max_{(T, \boldsymbol{\theta})} |\widehat{\mathcal{J}}(T; \boldsymbol{\theta}) - \mathcal{J}(T; \boldsymbol{\theta})| < \frac{\Delta}{2} \right\}$. On $\mathcal{E}_N$, for any $(T, \boldsymbol{\theta}) \neq (T^\star, \boldsymbol{\theta}^\star)$
\begin{align*}
    \widehat{\mathcal{J}}(T; \boldsymbol{\theta}) &> \mathcal{J}(T; \boldsymbol{\theta}) - \frac{\Delta}{2} \\
    &\ge \mathcal{J}(T^\star; \boldsymbol{\theta}^\star) + \Delta - \frac{\Delta}{2} \\
    &= \mathcal{J}(T^\star; \boldsymbol{\theta}^\star) + \frac{\Delta}{2} > \widehat{\mathcal{J}}(T^\star; \boldsymbol{\theta}^\star).
\end{align*}
Consequently, on $\mathcal{E}_N$, the empirical argmin must select $\widehat{T}_N = T^\star$. Because $\mathbb{P}(\mathcal{E}_N) \to 1$ as $N \to \infty$, it follows that $\mathbb{P}(\widehat{T}_N = T^\star) \to 1$.
\end{proof}

\section{Optimization Dynamics, Adaptive Topological Screening, and Restart Analysis}
\label{ec:sec:restarts}

This section provides the formal derivations and algorithmic specifications for the optimization layer introduced in Section~\ref{sec:CMA-ES-optimization} of the main text, covering the independent restart bounds, the mathematical rules governing adaptive regime detection, and the bounded continuous surrogate domain $\widetilde{\Theta}$.

\subsection{Independent Restarts and Basin Convergence Bound}
\label{ec:sec:restart-proof}

\begin{proposition}[Independent Restarts Success Bound]
\label{prop:ec-restarts}
Let $q_b \in (0, 1]$ denote the probability that a single run of CMA-ES with an evaluation budget of $b = 300$ locates an $\epsilon$-optimal basin of the DEFI objective. Under algorithmic independence enforced by unshared pseudo-random seeds and independent initial distribution means $\mathbf{m}^{(0)}$, the failure probability across $r$ independent restarts satisfies
\begin{equation}
    \mathbb{P}(\text{All } r \text{ restarts fail}) = \prod_{j=1}^{r} \mathbb{P}(\text{Restart } j \text{ fails}) = (1 - q_b)^r \le e^{-r q_b}.
    \label{eq:ec-restarts-bound}
\end{equation}
Furthermore, the number of independent restarts required to achieve the first basin discovery follows a geometric distribution with expectation $\mathbb{E}[R] = 1/q_b$, incurring an expected computational expenditure of $b / q_b$ objective evaluations.
\end{proposition}

\begin{proof}
Because pseudo-random seeds are disjoint and initial sample populations do not share covariance evolution paths, basin discovery events $\{E_j\}_{j=1}^r$ are mutually independent Bernoulli trials with parameter $q_b$. The joint failure probability is $(1 - q_b)^r$. Applying the elementary inequality $1 - x \le e^{-x}$ for all $x \in \mathbb{R}$ with $x = q_b$ yields $(1 - q_b)^r \le (e^{-q_b})^r = e^{-r q_b}$. Standard properties of the geometric distribution yield the expected restart count $\mathbb{E}[R] = 1/q_b$ and expected evaluation cost $b / q_b$. 
\end{proof}

\subsection{Adaptive Topological Screening and Regime Classification}
\label{ec:sec:regime-screening}

To prevent the optimizer from allocating evaluation budget to geometrically incompatible covariance parameterizations, an exploratory pre-conditioning routine analyzes the sample point-cloud before CMA-ES begins.

\paragraph{Density Contrast Estimation.}
Given the pairwise Euclidean distance matrix $\mathbf{D} \in \mathbb{R}^{N \times N}$, we evaluate the $k$-nearest-neighbor ($k$-NN) distance for each observation at scale $k_{\mathrm{nn}} = \max(3, \operatorname{round}(0.02 N))$
\begin{equation}
    r_i^{(k)} = \operatorname{sort}(\{D_{ij} : j \ne i\})_{k_{\mathrm{nn}}}, \qquad \hat{\rho}_i = \frac{1}{(r_i^{(k)})^P + 10^{-10}}.
    \label{eq:ec-knn-dens}
\end{equation}
The global empirical density contrast ratio is defined by the 95th-to-5th percentile spread:
\begin{equation}
    \kappa_{\mathrm{dens}} = \frac{Q_{0.95}(\{\hat{\rho}_i\})}{Q_{0.05}(\{\hat{\rho}_i\})}.
    \label{eq:ec-contrast-ratio}
\end{equation}
Additionally, a preliminary, unregularized Gaussian mixture fit is executed across $K \in \{1, \dots, \min(8, K_{\max})\}$ using standard EM to extract baseline estimates of component count $\widehat{K}_{\mathrm{BIC}}$ and dominant parameterization family $m_{\mathrm{BIC}}$.

\paragraph{Regime Classification Rules.}
\label{ec:RCR}
Let $\mathcal{M}_{\mathrm{asym}} = \{\texttt{VII}, \texttt{VEI}, \texttt{VVI}, \texttt{VEE}, \texttt{VVE}, \texttt{VEV}, \texttt{VVV}\}$ denote variable-volume covariance families and $\mathcal{M}_{\mathrm{eq}} = \{\texttt{EII}, \dots, \texttt{VVV}\}$ denote the full pool including equal-volume models, Datasets are assigned to one of three pre-conditioning regimes:

\vspace{2mm}

\begin{adjustbox}{max width=\textwidth, center}
\scriptsize
\renewcommand{\arraystretch}{0.95}
\setlength{\tabcolsep}{3.5pt}
\begin{tabular}{@{}l l c c l@{}}
\toprule
\textbf{Structural Regime} & \textbf{Activation Condition} & \textbf{Pool} & $\boldsymbol{\tilde{K}_0}$ & \textbf{Penalty Calibration ($\bar{\gamma}_2, \, \gamma_{2,0}$)} \\
\midrule
Asymmetric Manifold & $(\kappa_{\mathrm{dens}} > 20.0 \land N > 200) \lor (N \ge 700 \land \widehat{K}_{\mathrm{BIC}} \ge 7)$ & $\mathcal{M}_{\mathrm{asym}}$ & $2.0$ & $\bar{\gamma}_2 = \max(2.0, \operatorname{round}(0.4N/\log N))$, $\gamma_{2,0} = 0.75\bar{\gamma}_2$ \\
Equal-Volume Manifold & $N \ge 700 \land m_{\mathrm{BIC}} \in \{\texttt{EEE}, \texttt{EEV}, \texttt{EII}\} \land \kappa_{\mathrm{dens}} < 15.0$ & $\mathcal{M}_{\mathrm{eq}}$ & $3.5$ & $\bar{\gamma}_2 = 2.0$, $\gamma_{2,0} = 1.0$ \\
Standard Heterogeneous & All remaining configurations & $\mathcal{M}_{\mathrm{asym}}$ & $3.0^*$ & $\bar{\gamma}_2 = \max(2.0, \operatorname{round}(0.2N/\log N))$, $\gamma_{2,0} = 1.0$ \\
\bottomrule
\multicolumn{5}{@{}l}{\scriptsize $^*$Initialized to $\tilde{K}_0 = 2.0$ for binary-split configurations.}
\end{tabular}
\end{adjustbox}

\subsection{Parameter Domain and Surrogate Search Space Encoding}
\label{ec:sec:search-space-encoding}

Table~\ref{tab:ec-cmaes-parameters} details the 15 continuous surrogate parameters comprising $\widetilde{\Theta} \subset \mathbb{R}^{15}$, their decoding functions, and their admissible box constraints $[\underline{\theta}_i, \bar{\theta}_i]$.

\begin{table}[!htbp] 
\centering
\footnotesize
\caption{Continuous surrogate parameter encoding for CMA-ES optimization ($d = 15$).}
\renewcommand{\arraystretch}{0.88}
\label{tab:ec-cmaes-parameters}
\begin{tabular}{c l c c l}
\toprule
\textbf{Index} & \textbf{Parameter Description} & \textbf{Lower ($\underline{\theta}$)} & \textbf{Upper ($\bar{\theta}$)} & \textbf{Decoding / Operational Mapping} \\
\midrule
1  & Relaxed Cluster Count $\tilde{K}$      & $2.0$  & $K_{\max}$              & Homotopy weights $w_k = \tilde{K} - \lfloor \tilde{K} \rfloor$; Eq.~\eqref{eq:smooth_k_interpolation} \\
2  & Covariance Family Coordinate $\tilde{m}$ & $1.0$  & $\vert{}\mathcal{M}\vert{}$         & Nearest-integer projection: $m = \mathcal{M}[\operatorname{round}(\tilde{m})]$ \\
3  & Sample-Size Penalty Weight $\gamma_1$   & $0.01$ & $1.0$                   & Directly scales $\Omega_N(K) = K \log N$ \\
4  & Dimension Penalty Weight $\gamma_2$     & $0.01$ & $\bar{\gamma}_2(N)$     & Directly scales $\Omega_P(K) = K \log P$; regime-calibrated \\
5  & Spectral Weight $a_1$                   & $0.10$ & $1.0$                   & Directly scales spectral entropy difference $\Delta H_\alpha(T)$ \\
6  & Partition Entropy Weight $a_2$          & $0.10$ & $1.0$                   & Directly scales empirical partition entropy $\widehat{H}_N(T)$ \\
7  & R\'enyi Entropy Order $\alpha$          & $0.10$ & $2.0$                   & Matrix R\'enyi order for graph state $\boldsymbol{\rho}$ \\
8  & Multiscale Bandwidth Scale $\sigma_{\mathrm{rel}}$ & $0.20$ & $3.0$        & Bandwidth scaling: $\sigma = \sigma_{\mathrm{rel}} \cdot \operatorname{median}(\mathbf{D}_{ij} > 0)$ \\
9  & Entropy Evaluation Family               & $1.0$  & $2.0$                   & Discrete family switch (R\'enyi vs.\ von Neumann limit) \\
10 & Graph Construction Family               & $1.0$  & $3.0$                   & Kernel weighting topology (Gaussian, multiscale, $k$-NN) \\
11 & Sparsification Neighborhood $k_{\mathrm{graph}}$ & $5$ & $30$             & Nearest-neighbor retention threshold for affinity graph \\
12 & Relevance Weight $b$                    & $0.00$ & $2.0$                   & Information bottleneck anchor weight $-b \widehat{I}_N(T; Y)$ \\
13 & Multiscale Mixture Mixing Weight        & $0.05$ & $0.90$                  & Convex combination weight for affinity multiscale kernel \\
14 & Balance Penalty Weight $\gamma_{\mathrm{bal}}$ & $0.15$ & $1.00$           & Floor-constrained weight on $\Omega_{\mathrm{floor}}(T; \tau)$ \\
15 & Balance Metric Formulation Mode         & $1.0$  & $3.0$                   & Squared hinge floor vs.\ relative entropy deficit \\
\bottomrule
\end{tabular}
\end{table}

All runs initialize the step-size parameter to $\sigma_0 = 0.20$ and anchor the initial mean $\mathbf{m}^{(0)}$ to mid-range coordinates, with the covariance coordinate $m_2^{(0)}$ anchored to the unconstrained elliptical family (\texttt{VEV}) whenever admissible. Objective evaluations mapping to degenerate candidate partitions containing splinter clusters below the cardinality constraint ($n_c < 2$) immediately abort and return the infeasibility penalty $P_{\mathrm{inf}} = +10^9$, preserving well-behaved evolution paths along feasible domain boundaries.

\section{Computational Invariance of Geodesic Embeddings}
\label{ec:sec:isomap-invariance}
\begin{proof}{Proof of Proposition 3 (Nested Subspace Invariance)}
\label{ec:proof:isomap-invariance}
Let $\mathbf{D}_{\mathrm{geo}} \in \mathbb{R}^{N \times N}$ be the shortest-path geodesic matrix and $\mathbf{B} = -\frac{1}{2}\mathbf{H}\mathbf{D}_{\mathrm{geo}}^2\mathbf{H}$ the double-centered Gram matrix, where $\mathbf{H} = \mathbf{I}_N - \frac{1}{N}\mathbf{1}\mathbf{1}^T$. Let $\mathbf{B} = \mathbf{U}\boldsymbol{\Lambda}\mathbf{U}^T$ be its spectral decomposition, with eigenvalues $\lambda_1 \ge \dots \ge \lambda_N$ and orthonormal eigenvectors $\mathbf{U}^T \mathbf{U} = \mathbf{I}_N$.

For a truncated rank $m \le M$, classical MDS coordinates are $\mathbf{Z}_m = \mathbf{U}_m \boldsymbol{\Lambda}_m^{1/2}$, where $\mathbf{U}_m = [\mathbf{u}_1, \dots, \mathbf{u}_m]$ and $\boldsymbol{\Lambda}_m = \operatorname{diag}(\lambda_1, \dots, \lambda_m)$. Evaluating the inner product of the coordinate matrix
\begin{equation*}
    \mathbf{Z}_m^T \mathbf{Z}_m = \left(\mathbf{U}_m \boldsymbol{\Lambda}_m^{1/2}\right)^T \left(\mathbf{U}_m \boldsymbol{\Lambda}_m^{1/2}\right) = \boldsymbol{\Lambda}_m^{1/2} \left(\mathbf{U}_m^T \mathbf{U}_m\right) \boldsymbol{\Lambda}_m^{1/2} = \boldsymbol{\Lambda}_m^{1/2} \mathbf{I}_m \boldsymbol{\Lambda}_m^{1/2} = \boldsymbol{\Lambda}_m.
\end{equation*}
Because $\boldsymbol{\Lambda}_m$ is strictly diagonal, its inverse is $\boldsymbol{\Lambda}_m^{-1} = \operatorname{diag}(\lambda_1^{-1}, \dots, \lambda_m^{-1})$. Substituting this directly into the Ordinary Least Squares dual projection matrix
\begin{equation*}
    \mathbf{V}_m = \mathbf{X}^T \mathbf{Z}_m (\mathbf{Z}_m^T \mathbf{Z}_m)^{-1} = \mathbf{X}^T \left(\mathbf{U}_m \boldsymbol{\Lambda}_m^{1/2}\right) \boldsymbol{\Lambda}_m^{-1} = \mathbf{X}^T \mathbf{U}_m \boldsymbol{\Lambda}_m^{-1/2}.
\end{equation*}
Expanding this product into individual columns
\begin{equation*}
    \mathbf{V}_m = \Big[\,\mathbf{v}_1 \;\;\cdots\;\; \mathbf{v}_m\,\Big], \qquad \text{where} \quad \mathbf{v}_j = \frac{1}{\sqrt{\lambda_j}} \mathbf{X}^T \mathbf{u}_j.
\end{equation*}
Each column $\mathbf{v}_j \in \mathbb{R}^P$ depends solely on the $j$-th eigenpair $(\lambda_j, \mathbf{u}_j)$ and is completely invariant to candidate dimension expansions $m \to m+1$.
\end{proof}

\begin{remark}[Spectral Tail Concentration under R\'enyi Orders $\alpha \ge 2$]
\label{ec:rem:spectral-tail}
Parallel computational efficiency applies to the graph state $\boldsymbol{\rho} = \mathbf{L} / \operatorname{Tr}(\mathbf{L})$. For any R\'enyi order $\alpha \ge 2$, the spectral power sum $\operatorname{Tr}(\boldsymbol{\rho}^\alpha) = \sum_{i=1}^N \lambda_i^\alpha$ decays rapidly along the sorted spectrum $\lambda_1 \ge \lambda_2 \ge \dots \ge \lambda_N \ge 0$. Truncating the calculation at the leading $k \ll N$ eigenvalues via partial Lanczos iteration (e.g., using implicit restarts) incurs a residual truncation error bounded by
\begin{equation}
    \sum_{i=k+1}^N \lambda_i^\alpha \;\le\; (N - k)\,\lambda_{k+1}^\alpha.
\end{equation}
For $\alpha = 2$, this bound guarantees that evaluating $\Delta H_2(T)$ requires only the active principal components of the graph Laplacian, avoiding full $\mathcal{O}(N^3)$ decompositions during repeated fitness evaluations.
\end{remark}

\begin{proposition}[Per-Iteration Complexity of Manifold Rank Selection]
\label{prop:complexity}
Let $N$ be the sample size, $P$ the manifest feature count, and $M = m_{\max} - 1$ the candidate dimensions evaluated. Whereas classical Isomap evaluates candidate ranks via sample-distance correlations requiring $\mathcal{O}(M N^2)$ operations, DEFI-DR computes the double-centered Gram decomposition once in $\mathcal{O}(m_{\max} N^2)$ and evaluates candidate ranks via dual feature-space projections $\mathbf{V}_m$ and $P \times P$ correlation matrices, scaling as $\mathcal{O}(M(N P m_{\max} + P^3))$. In regimes where $N \gg P$, this avoids the quadratic sample-pair bottleneck $\mathcal{O}(M N^2)$.
\end{proposition}

\section{Software Setup}
\label{ec:sec:software}

Data generation and analysis were executed in R 4.6.1 (\texttt{aarch64-apple-darwin25.4.0}) with \texttt{future.seed = TRUE}) and single-threaded BLAS operations (\texttt{RhpcBLASctl} v0.23.42). Population generation and item categorization were conducted using \texttt{latentFactoR} (v0.0.7). Exploratory network models and dimensionalities were assessed using \texttt{EGANet} (v2.4.1) and \texttt{psych} (v2.6.5), while SEM fit was verified via \texttt{lavaan} (v0.6.21). Sobol global sensitivity metrics ($S_1, S_T$) were generated using \texttt{sensitivity} (v1.31.0) and \texttt{randtoolbox} (v2.0.5). Simulations were parallelized across grid conditions using \texttt{future} (v1.70.0) and \texttt{future.apply} (v1.20.2). Data wrangling was performed using \texttt{tidyverse} (v2.0.0).

\section{Supplementary Tables and Figures}
\label{ectables}

\textit{Note on Table~\ref{tab:oracle-recoveryMC} vs.\ Table~\ref{tab:oracle-recoveryPAFull}:} Table~2 in the main manuscript and Table~\ref{tab:oracle-recoveryPAFull} in this companion report exact recovery evaluated on the asymptotic population-analytic covariance matrices, isolating the criterion's deterministic ranking properties. Table~\ref{tab:oracle-recoveryMC} reports the corresponding finite-sample Monte Carlo simulation ($R = 100$ independent synthetic realizations per cell), where empirical sampling error and finite-sample variance introduce replication noise.

\begin{table}[ht]
\centering
\tiny
\caption{Study 3: Estimated cluster count using DEFI-HC ($\widehat{K}$) and ARI performance across benchmark datasets ($K^{\dagger}$ denotes true reference count).}
\label{tab:clustering-results-hc}
\begin{adjustbox}{width=0.7\textwidth}
\begin{tabular}{lc*{4}{cc}}
\toprule
\multirow{2}{*}{Dataset} & \multirow{2}{*}{True $K^{\dagger}$} & \multicolumn{2}{c}{$k$-means (Sil)} & \multicolumn{2}{c}{GMM (BIC)} & \multicolumn{2}{c}{HC-Ward (Sil)} & \multicolumn{2}{c}{\textbf{DEFI-HC}} \\
\cmidrule(lr){3-4} \cmidrule(lr){5-6} \cmidrule(lr){7-8} \cmidrule(lr){9-10}
 & & $\widehat{K}$ & ARI & $\widehat{K}$ & ARI & $\widehat{K}$ & ARI & $\widehat{K}$ & ARI \\
\midrule
Iris         & 3 & 2 & 0.5681 & 2 & 0.5681 & 2 & 0.5438 & \textbf{3} & \textbf{0.7430} \\
Seeds        & 3 & 2 & 0.4805 & 4 & 0.5795 & 2 & 0.4762 & \textbf{3} & \textbf{0.6863} \\
WingNut      & 2 & 2 & 0.8595 & 6 & 0.2771 & 2 & 0.6011 & 3 & \textbf{0.8825} \\
Glass        & 6 & 2 & 0.1930 & 5 & 0.1470 & 4 & 0.1732 & 5 & \textbf{0.2902} \\
Ionosphere   & 2 & 4 & 0.2860 & 2 & 0.2519 & 4 & 0.3171 & 6 & \textbf{0.3502} \\
Wine         & 3 & 3 & 0.8975 & 3 & \textbf{0.9297} & 3 & 0.7899 & 4 & 0.6213 \\
BreastCancer & 2 & 2 & \textbf{0.6707} & 6 & 0.1388 & 2 & 0.5750 & 2 & 0.6163 \\
Banknote     & 2 & 5 & 0.2112 & 5 & \textbf{0.3926} & 5 & 0.3308 & 5 & 0.3308 \\
\midrule
\textbf{Mean} & -- & 2.62 & 0.5208 & 4.12 & 0.4106 & 3.00 & 0.4509 & 3.88 & \textbf{0.5651} \\
\bottomrule
\end{tabular}
\end{adjustbox}
\end{table}

\begin{table}[htbp]
\centering
\scriptsize
\renewcommand{\arraystretch}{0.8}
\caption{Complete real-world benchmark suite (26 UCI datasets and MNIST digits) for intrinsic dimensionality evaluation.}
\label{ec:tab:study4-datasets}
\begin{tabularx}{\textwidth}{@{}r l c c c | r l c c c@{}}
\toprule
\textbf{\#} & \textbf{Dataset} & $\boldsymbol{p}$ & $\boldsymbol{n}$ & $\boldsymbol{m^*}$ & 
\textbf{\#} & \textbf{Dataset} & $\boldsymbol{p}$ & $\boldsymbol{n}$ & $\boldsymbol{m^*}$ \\
\midrule
1  & Iris            & 4  & 150  & 2  & 15 & User Knowledge   & 5  & 400  & 2  \\
2  & Wine            & 13 & 178  & 3  & 16 & Spambase        & 57 & 500  & 8  \\
3  & Vehicle         & 18 & 846  & 4  & 17 & SyntheticControl& 60 & 500  & 6  \\
4  & Ecoli           & 7  & 336  & 3  & 18 & StatlogSat      & 36 & 500  & 4  \\
5  & Breast Cancer   & 30 & 569  & 2  & 19 & Thyroid         & 5  & 215  & 2  \\
6  & Seeds           & 7  & 210  & 3  & 20 & BloodTransfusion& 4  & 748  & 2  \\
7  & Haberman        & 3  & 306  & 2  & 21 & AuditRisk       & 18 & 500  & 4  \\
8  & Banknote        & 4  & 1372 & 2  & 22 & CreditDefault   & 23 & 500  & 5  \\
9  & WineQuality     & 11 & 1599 & 4  & 23 & SteelPlates     & 27 & 500  & 5  \\
10 & Mammographic    & 5  & 830  & 2  & 24 & VehicleSilh     & 18 & 500  & 4  \\
11 & Glass           & 9  & 214  & 4  & 25 & CarEval         & 6  & 500  & 3  \\
12 & Diabetes        & 8  & 500  & 3  & 26 & WirelessIndoor  & 7  & 500  & 3  \\
13 & DryBean         & 16 & 500  & 5  & 27 & MNIST Digits    & 64 & 500  & 10 \\
14 & Rice            & 7  & 500  & 3  &    &                 &    &      &    \\
\bottomrule
\end{tabularx}
\end{table}

\begin{table}[htbp]
\centering
\tiny
\setlength{\tabcolsep}{3.2pt} 
\caption{Study 1: Monte Carlo-Based Comprehensive exact recovery and dimension retention rates (\%) across latent factor structures ($R=100$ replications per cell).}
\label{tab:oracle-recoveryMC}
\begin{adjustbox}{width=\textwidth}
\renewcommand{\arraystretch}{0.8}
\begin{tabular}{ll c ccc cc ccc}
\toprule
 & & \textbf{Proposed} & \multicolumn{3}{c}{\textbf{Network \& Retention Baselines}} & \multicolumn{2}{c}{\textbf{Entropic Criteria}} & \multicolumn{3}{c}{\textbf{CFA Fit Index Rules}} \\
\cmidrule(lr){3-3} \cmidrule(lr){4-6} \cmidrule(lr){7-8} \cmidrule(lr){9-11}
\textbf{Simulation Axis} & \textbf{Level} & \textbf{DEFI-FA} & \textbf{EGA-TMFG} & \textbf{PA} & \textbf{Kaiser} & \textbf{TEFI.vn} & \textbf{Lap.-BIC} & \textbf{CFI} & \textbf{TLI/RMSEA} & \textbf{SRMR} \\
\midrule
\multirow{4}{*}{Factor correlation ($r$)}
 & 0.3 (Weak) & 92.6 & 99.9  & \textbf{100.0} & 96.4 & 66.2 & 7.0 & 76.6 & 68.5 & 54.8 \\
 & 0.5 (Med) & 94.6 & 93.6 & \textbf{99.7} & 96.2 & 73.5 & 2.8 & 66.3 & 60.9 & 32.1 \\
 & 0.7 (High) & 95.9 & 77.0 & \textbf{97.2} & 95.9 & 35.3 & 0.3 & 53.8 & 46.4 & 10.6 \\
 & 0.9 (V. High*) & 95.9 & 77.0 & \textbf{97.2} & 95.9 & 35.3 & 0.3 & 53.8 & 46.4 & 10.6 \\
\midrule
\multirow{4}{*}{Factor loading ($\lambda$)}
 & 0.30 (Weak*)  & \textbf{99.4} & 82.2 & 96.8  & 88.5 & 51.2 & 0.0 & 0.0 & 0.0  & 19.5 \\
 & 0.50 (Med)  & \textbf{99.4} & 82.2 & 96.8  & 88.5 & 51.2 & 0.1 & 10.1 & 1.8  & 19.5 \\
 & 0.70 (High)  & \textbf{100.0} & 94.1 & \textbf{100.0} &\textbf{ 100.0} & 69.7 & 0.1 & 86.6 & 74.0 & 61.4 \\
 & 0.90 (V. High) & 93.5 & 77.0 & \textbf{100.0} & \textbf{100.0}& 54.1 & 9.9 & \textbf{100.0} & \textbf{100.0} & 16.6 \\
\midrule
\multirow{3}{*}{Measurement scale}
 & Binary       & \textbf{99.5} & 89.2  & 97.9  & 89.4  & 57.4 & 0.2 & 55.0 & 49.8 & 36.7 \\
 & Ordinal (5-pt)& 97.2 & 92.5 & \textbf{99.9} & 99.6  & 60.2 & 1.9 & 68.2 & 60.1 & 31.6 \\
 & Continuous   & 86.2 & 91.4 & 99.0  & \textbf{99.5} & 57.4 & 8.0 & 73.5 & 65.9 & 29.1 \\
\bottomrule
\end{tabular}
\end{adjustbox}
\end{table}
\begin{table}[ht]
\tiny
\centering
\caption{Study 1: Full (including skew) Population Analytic-based exact recovery and dimension retention rates (\%) across latent factor structures. Methods are grouped by paradigm: network/retention benchmarks, entropic criteria, and confirmatory fit indices. Best performing rate per condition in bold.}
\label{tab:oracle-recoveryPAFull}
\renewcommand{\arraystretch}{0.8}
\begin{adjustbox}{width=\textwidth}
\begin{tabular}{ll c ccc cc ccc}
\toprule
 & & \textbf{Proposed} & \multicolumn{3}{c}{\textbf{Network \& Retention Baselines}} & \multicolumn{2}{c}{\textbf{Entropic Criteria}} & \multicolumn{3}{c}{\textbf{CFA Fit Index Rules}} \\
\cmidrule(lr){3-3} \cmidrule(lr){4-6} \cmidrule(lr){7-8} \cmidrule(lr){9-11}
\textbf{Simulation Axis} & \textbf{Level} & \textbf{DEFI-FA} & \textbf{EGA-TMFG} & \textbf{PA} & \textbf{Kaiser} & \textbf{TEFI.vn} & \textbf{Lap.-BIC} & \textbf{CFI} & \textbf{TLI/RMSEA} & \textbf{SRMR} \\
\midrule
\multirow{4}{*}{Factor correlation ($r$)}
 & 0.3 (Weak)     & 94.4           & \textbf{100.0} & 75.0 & \textbf{100.0} & 66.7 & 16.7 & 66.7          & 58.3 & 44.4 \\
 & 0.5 (Med)      & 94.4           & \textbf{100.0} & 75.0 & \textbf{100.0} & 55.6 & 22.2 & 50.0          & 50.0 & 41.7 \\
 & 0.7 (High)     & \textbf{100.0} & \textbf{100.0} & 55.6 & \textbf{100.0} & 41.7 & 19.4 & 41.7          & 41.7 & 38.9 \\
 & 0.9 (V. High) & \textbf{100.0} & 0.0            & 5.6  & 0.0            & 19.4 & 2.8  & 22.2          & 16.7 & 25.0 \\
\midrule
\multirow{4}{*}{Factor loading ($\lambda$)}
 & 0.30 (Weak)   & \textbf{100.0} & 75.0           & 11.1 & 75.0           & 19.4 & 0.0  & 0.0           & 0.0  & 0.0  \\
 & 0.50 (Med)     & \textbf{100.0} & 75.0           & 50.0 & 75.0           & 72.2 & 0.0  & 16.7          & 8.3  & 11.1 \\
 & 0.70 (High)    & \textbf{100.0} & 75.0           & 75.0 & 75.0           & 58.3 & 0.0  & 66.7          & 66.7 & 55.6 \\
 & 0.90 (V. High) & 88.9           & 75.0           & 75.0 & 75.0           & 33.3 & 61.1 & \textbf{97.2} & 91.7 & 83.3 \\
\midrule
\multirow{3}{*}{Measurement scale}
 & Binary         & \textbf{100.0} & 75.0           & 50.0 & 75.0           & 47.9 & 8.3  & 35.4          & 31.2 & 35.4 \\
 & Ordinal (5-pt) & \textbf{100.0} & 75.0           & 50.0 & 75.0           & 41.7 & 16.7 & 50.0          & 43.8 & 37.5 \\
 & Continuous     & \textbf{91.7}  & 75.0           & 58.3 & 75.0           & 47.9 & 20.8 & 50.0          & 50.0 & 39.6 \\
\midrule
\multirow{3}{*}{Skewness}
 & $-2$           & \textbf{97.2}  & 75.0           & 52.8 & 75.0           & 45.8 & 15.3 & 45.1          & 41.7 & 37.5 \\
 & $0$            & \textbf{97.2}  & 75.0           & 52.8 & 75.0           & 45.8 & 15.3 & 45.1          & 41.7 & 37.5 \\
 & $2$            & \textbf{97.2}  & 75.0           & 52.8 & 75.0           & 45.8 & 15.3 & 45.1          & 41.7 & 37.5 \\
\midrule
\multirow{3}{*}{Kurtosis}
 & Heavy          & \textbf{100.0} & 75.0           & 50.0 & 75.0           & 45.8 & 12.5 & 43.8          & 41.7 & 37.5 \\
 & Light          & \textbf{95.8}  & 75.0           & 54.2 & 75.0           & 45.8 & 16.7 & 45.8          & 41.7 & 37.5 \\
 & Mesokurtic     & \textbf{95.8}  & 75.0           & 54.2 & 75.0           & 45.8 & 16.7 & 45.8          & 41.7 & 37.5 \\
\bottomrule
\end{tabular}
\end{adjustbox}
\end{table}

\begin{table}[htbp]
\centering
\tiny
\setlength{\tabcolsep}{3.5pt} 
\caption{Mean performance metrics across dimensionality reduction methods over 26 benchmark datasets. Metrics are grouped by dimensionality selection accuracy, manifold reconstruction fidelity, and correlation preservation. Best non-overextracted values are in bold.}
\label{tab:method_comparison}
\begin{adjustbox}{width=\textwidth}
\renewcommand{\arraystretch}{0.88}
\begin{tabular}{l ccc cccc ccccc}
\toprule
 & \multicolumn{3}{c}{\textbf{Dimension Retention}} & \multicolumn{4}{c}{\textbf{Manifold \& Reconstruction}} & \multicolumn{5}{c}{\textbf{Correlation \& Information Criteria}} \\
\cmidrule(lr){2-4} \cmidrule(lr){5-8} \cmidrule(lr){9-13}
\textbf{Method} 
& \shortstack{$|\Delta m|$\\Over} 
& \shortstack{Prop.\\Err.} 
& \shortstack{Eff.\\$\widehat{m}$} 
& \shortstack{Geo.\\Stress} 
& \shortstack{MAE} 
& \shortstack{Var.\\(\%)} 
& \shortstack{Frob.\\$\|\mathbf{E}\|_F$} 
& \shortstack{Corr.\\Pres.} 
& \shortstack{Corr.\\RMSE} 
& \shortstack{SNR} 
& \shortstack{Raw BIC} 
& \shortstack{Raw AIC} \\
\midrule
VAE               & 1.31  & 0.383 & 2.90 & 0.2599 & 0.1330 & 77.6 & 0.2967 & 0.7217 & 0.3108 & 206.35 & -66.5   & -80.0   \\
\textbf{DEFI-DR}  & \textbf{0.04} & \textbf{0.013} & 3.18 & \textbf{0.0668} & 0.1088 & \textbf{97.6} & \textbf{0.2259} & \textbf{0.9689} & \textbf{0.0394} & \textbf{270.33} & -255.8  & -270.8  \\
Isomap            & 14.00 & 3.104 & 3.29 & 0.0666 & 0.0547 & 99.7 & 0.3106 & 0.9761 & 0.0303 & 204.55 & -285.2  & -358.3  \\
Kaiser            & 0.38  & 0.162 & 3.10 & 0.1181 & 0.1425 & 83.8 & 0.3384 & 0.8448 & 0.1818 & 205.57 & -111.7  & -126.8  \\
Kernel PCA (RBF)  & 8.36  & 1.735 & 3.65 & 0.1630 & 0.1001 & 95.7 & 0.1463 & 0.9606 & 0.0523 & 154.18 & -1081.7 & -1132.8 \\
LDA               & 1.31  & 0.309 & 2.43 & 0.2166 & 0.2876 & 80.0 & 0.4013 & 0.8449 & 0.2053 & 39.40  & 89.3    & 77.0    \\
Parallel Analysis & 0.42  & 0.176 & 3.06 & 0.1181 & 0.1479 & 85.5 & 0.3119 & 0.8506 & 0.1769 & 229.39 & -215.0  & -229.8  \\
Robust PCA (MCD)  & 0.88  & 0.253 & 2.87 & 0.1916 & 0.2656 & 84.5 & 0.3765 & 0.8679 & 0.1794 & 33.26  & 198.0   & 184.2   \\
Sparse PCA        & 13.15 & 2.913 & 3.77 & 0.1724 & 0.0533 & 98.3 & 0.0808 & 0.9740 & 0.0321 & 594.81 & -2201.3 & -2271.9 \\
TEFI.vn           & 1.38  & 0.404 & 2.61 & 0.2224 & 0.2853 & 77.4 & 0.4038 & 0.7515 & 0.2874 & 104.53 & 81.1    & 67.1    \\
Var $\ge 80\%$    & 1.04  & 0.285 & 2.99 & 0.1873 & 0.2586 & 86.6 & 0.3612 & 0.8838 & 0.1586 & 33.83  & 147.9   & 133.5   \\
\bottomrule
\end{tabular}
\end{adjustbox}
\end{table}

\begin{figure}[!ht]
\centering
\includegraphics[width=0.85\textwidth]{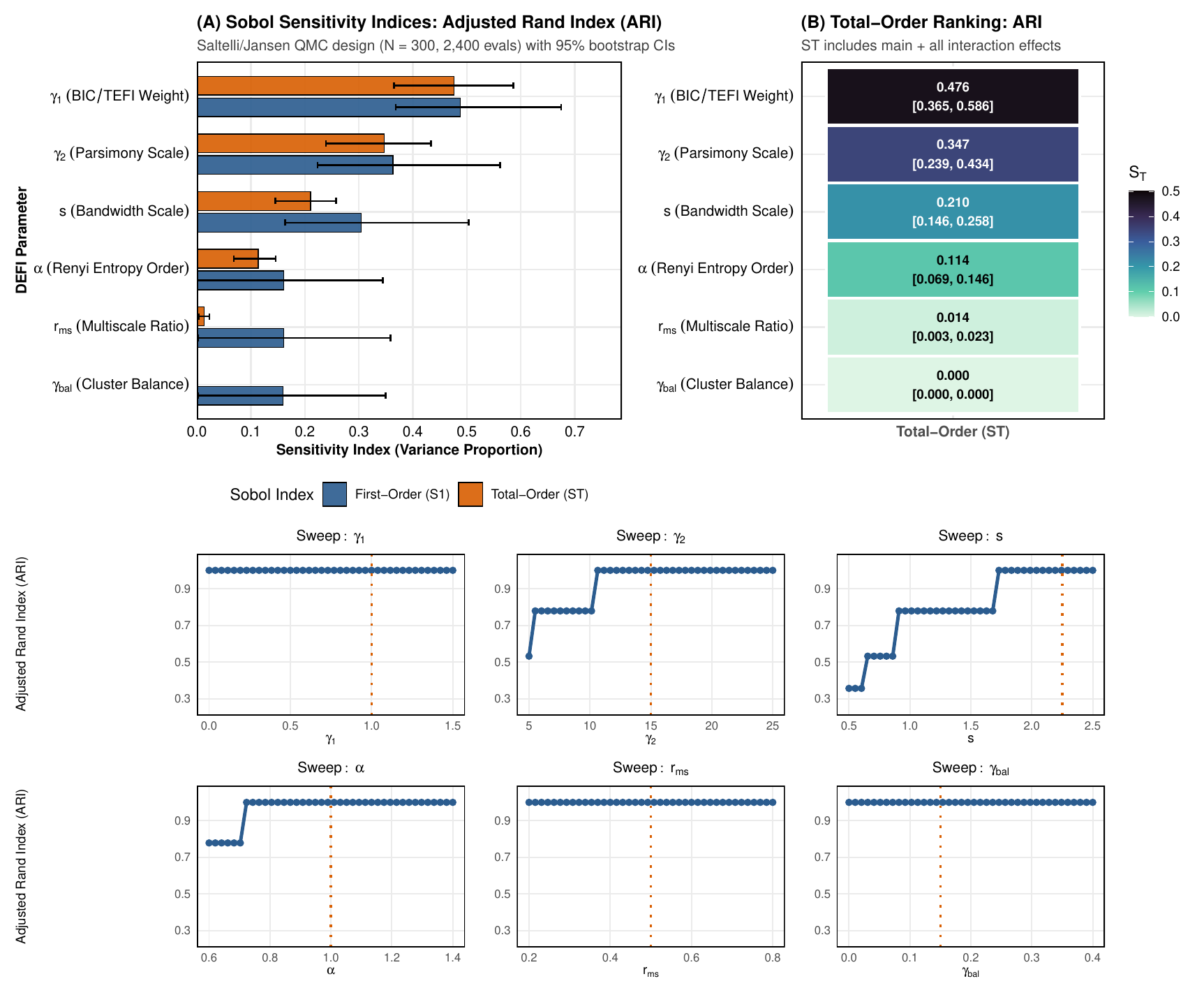}
\caption{Global Sobol sensitivity analysis for selected cluster count $\widehat{K}$ on \emph{TwoDiamonds} ($N = 300$, $2{,}400$ Saltelli-Jansen quasi-Monte Carlo evaluations, 95\% bootstrap CIs). \textbf{(A--B)} First-order ($S_1$) and total-order ($S_T$) variance decompositions across DEFI parameters. \textbf{Bottom:} Univariate parameter sweeps of $\widehat{K}$ around the default operating configuration (dotted vertical lines).}
\label{fig:ec-sobol-gsa}
\end{figure}

\clearpage

\section{Notation and Symbol Index} 
\label{ec:sec:notation} 
Table~\ref{tab:ec-notation} catalogs the mathematical symbols, parameter domains, and operational definitions used across the paper and this Electronic Companion. 

\begin{table}[!htbp] 
\centering 
\footnotesize 
\renewcommand{\arraystretch}{0.83} 
\caption{Consolidated index of mathematical notation, parameter domains, and definitions.} 
\label{tab:ec-notation} \begin{tabularx}{\textwidth}{@{}p{0.26\textwidth} X@{}} 
\toprule 
\textbf{Symbol / Notation} & \textbf{Operational Definition and Parameter Domain} 
\\ \midrule 
\multicolumn{2}{@{}l}{\textbf{Data Dimensions and Candidate Sets}} \\ $N, P, M$ & Sample size ($N$), feature count ($P$), and partitioned domain size $M \in \{N, P\}$. 
\\ $\mathbf{X}, \mathbf{R}_{\mathbf{X}}$ & Manifest data matrix $\mathbf{X} \in \mathbb{R}^{N \times P}$ and empirical correlation matrix $\mathbf{R}_{\mathbf{X}} \in \mathbb{R}^{P \times P}$. 
\\ $\mathcal{T}, \Theta, \widetilde{\Theta}$ & Candidate partition family $\mathcal{T}$, parameter domain $\Theta$, and surrogate space $\widetilde{\Theta} \subset \mathbb{R}^{15}$. \\ 
\midrule 
\multicolumn{2}{@{}l}{\textbf{Partitions and Information Bottleneck (IB)}} 
\\ $T, K, K_{\max}$ & Candidate assignment map $T: \{1,\dots,M\} \to \{1,\dots,K\}$, cluster count $K$, and search ceiling $K_{\max}$. \\ $n_c(T), \widehat{p}_T(c)$ & Community cardinality $n_c(T) = \sum_i \mathbb{I}(T_i = c)$ and empirical proportion $\widehat{p}_T(c) = n_c(T) / M$. 
\\ $\widehat{H}_M(T), \widetilde{H}_{\mathrm{size}}(T)$ & Empirical partition entropy $-\sum_c \widehat{p}_T(c) \log \widehat{p}_T(c)$ and normalized size entropy $\widehat{H}_M / \log K \in [0, 1]$. 
\\ $Y, \widehat{p}_{TY}, \widehat{I}_M(T; Y)$ & Self-supervised anchor $Y \in \{1,\dots,K_Y\}^M$, joint contingency $\widehat{p}_{TY}$, and plug-in mutual information. \\ \midrule \multicolumn{2}{@{}l}{\textbf{Affinity Graph State and Spectral Cuts}} \\ $\mathbf{W}, \mathbf{D}, \mathbf{L}$ & Affinity matrix ($\mathbf{W}$), degree matrix ($\mathbf{D}$), and combinatorial Laplacian $\mathbf{L} = \mathbf{D} - \mathbf{W}$. \\ $V, S$ & Graph volume $V = \operatorname{Tr}(\mathbf{L}) = \sum_i d_i$ and quadratic trace $S = \operatorname{Tr}(\mathbf{L}^2) = \sum_i d_i^2 + \sum_{i \ne j} W_{ij}^2$. \\ $\boldsymbol{\rho}, H_\alpha(\boldsymbol{\rho}), H_{\mathrm{vN}}$ & Normalized density state $\boldsymbol{\rho} = \mathbf{L}/V$, matrix R\'enyi entropy of order $\alpha$, and von Neumann limit ($\alpha \to 1$). \\ $\Delta H_\alpha(T), \widetilde{\boldsymbol{\rho}}_T$ & Signed spectral entropy difference $H_\alpha(\widetilde{\boldsymbol{\rho}}_T) - H_\alpha(\boldsymbol{\rho})$ evaluated on severed graph state $\widetilde{\boldsymbol{\rho}}_T$. \\ $c_i, C(T), \varepsilon(T)$ & Vertex cut incidence $c_i = \sum_{j: T_j \ne T_i} W_{ij}$, total cut volume $C(T)$, and quadratic trace penalty $\varepsilon(T)$. \\ \midrule \multicolumn{2}{@{}l}{\textbf{DEFI Objective and Regularization Weights}} \\ $\widehat{J}_{\mathrm{DEFI}}, \mathcal{J}_{\mathrm{RDIB}}$ & DEFI objective $\widehat{J}_{\mathrm{DEFI}}(T; \boldsymbol{\theta})$ and population regularized deterministic IB target. \\ $a_1, a_2, b$ & Nonnegative weights for spectral concentration ($a_1$), partition entropy ($a_2$), and relevance ($b$). \\ $\boldsymbol{\gamma}, \eta$ & Penalty weights for sample size ($\gamma_1$), dimension ($\gamma_2$), balance ($\gamma_{\mathrm{bal}}$), fragmentation ($\gamma_{\mathrm{frag}}$), and topology ($\eta$). \\ $\Omega_N, \Omega_P, \Omega_{\mathrm{floor}}$ & Dimension penalties ($K \log N, K \log P$) and squared hinge balance floor $[\max(0, \tau - \widetilde{H}_{\mathrm{size}}(T))]^2$. \\ $\Omega_{\mathrm{frag}}, \Omega_{\mathrm{graph}}$ & Fraction of splinter clusters ($n_c < n_{\min}$) and graph topological regularizer (average normalized cut). \\ \midrule \multicolumn{2}{@{}l}{\textbf{Continuous Relaxation and CMA-ES Optimization}} \\ $\tilde{K}, k_{\mathrm{fl}}, k_{\mathrm{ce}}, w_k$ & Relaxed cluster coordinate $\tilde{K} \in [2, K_{\max}]$, bounding integers $\lfloor \tilde{K} \rfloor, \lceil \tilde{K} \rceil$, and interpolation weight $w_k \in [0, 1)$. \\ $\tilde{m}, m_{\mathrm{cov}}, \mathcal{M}_{\mathrm{cache}}$ & Continuous covariance axis $\tilde{m}$, decoded family $m_{\mathrm{cov}} \in \mathcal{M}$, and offline in-memory EM lookup cache. \\ $\widetilde{J}_{\mathrm{DEFI}}, P_{\mathrm{inf}}$ & Piecewise linear continuous surrogate homotopy and infeasibility penalty ($+10^9$ for $n_c < 2$). \\ $\mathbf{m}^{(g)}, \mathbf{C}^{(g)}, \sigma_{\mathrm{step}}^{(g)}$ & CMA-ES sampling distribution mean vector, adaptive covariance matrix, and step size at generation $g$. \\ $\widehat{G}, J^\dagger, \widehat{J}, \epsilon$ & Oracle gap $\widehat{G} = \widehat{J} - J^\dagger$, reference value $J^\dagger$, incumbent $\widehat{J}$, and non-inferiority tolerance $\epsilon \in \{0.0, 0.05\}$. \\ \midrule \multicolumn{2}{@{}l}{\textbf{Manifold Representation Selection (DEFI-DR)}} \\ $\mathbf{D}_{\mathrm{geo}}, \mathbf{H}, \mathbf{B}$ & Geodesic shortest-path matrix, centering operator $\mathbf{H} = \mathbf{I} - \frac{1}{N}\mathbf{1}\mathbf{1}^T$, and Gram matrix $\mathbf{B} = -\frac{1}{2}\mathbf{H}\mathbf{D}_{\mathrm{geo}}^{\circ 2}\mathbf{H}$. \\ $\mathbf{Z}_m, \mathbf{V}_m, m^*$ & Truncated MDS coordinates $\mathbf{Z}_m$, dual projection $\mathbf{V}_m = \mathbf{X}^T \mathbf{U}_m \boldsymbol{\Lambda}_m^{-1/2}$, and selected intrinsic dimension $m^*$. \\ \midrule \multicolumn{2}{@{}l}{\textbf{Stochastic Block Model (SBM) Analysis}} \\ $q, s, p_{\mathrm{in}}, p_{\mathrm{out}}$ & Planted SBM community count ($q$), block size ($s$), and internal/external Bernoulli edge connection probabilities. \\ $\pi, \pi_n, B_\alpha(\epsilon), V_{P,n}$ & Internal-edge mass (population $\pi$, empirical $\pi_n$), entropy root $B_\alpha$, and CLT asymptotic variance $V_{P,n}$. \\ 
\bottomrule 
\end{tabularx}
 \end{table}

\end{document}